\input harvmac
\input amssym
\input epsf
\let\includefigures=\iftrue
%
% the following is to use blackboard bold fonts --
%\let\useblackboard=\iftrue
%
% activate this if you don't have them.
%\let\useblackboard=\iffalse
%
% You might also need to remove this line.
\newfam\black
\noblackbox
\includefigures
\message{If you do not have epsf.tex (to include figures),}
\message{change the option at the top of the tex file.}
\def\figin{\epsfcheck\figin}\def\figins{\epsfcheck\figins}
\def\epsfcheck{\ifx\epsfbox\UnDeFiNeD
\message{(NO epsf.tex, FIGURES WILL BE IGNORED)}
\gdef\figin##1{\vskip2in}\gdef\figins##1{\hskip.5in}% blank space instead
\else\message{(FIGURES WILL BE INCLUDED)}%
\gdef\figin##1{##1}\gdef\figins##1{##1}\fi}
\def\DefWarn#1{}

\def\figinsert{\goodbreak\midinsert}
\def\ifig#1#2#3{\DefWarn#1\xdef#1{fig.~\the\figno}
\writedef{#1\leftbracket fig.\noexpand~\the\figno}%
\figinsert\figin{\centerline{#3}}\medskip\centerline{\vbox{\baselineskip12pt
\advance\hsize by -1truein\noindent\footnotefont{\bf
Fig.~\the\figno:} #2}}
\bigskip\endinsert\global\advance\figno by1}
%%%
\else
\def\ifig#1#2#3{\xdef#1{fig.~\the\figno}
\writedef{#1\leftbracket fig.\noexpand~\the\figno}%
%\figinsert\figin{\centerline{#3}}\medskip\centerline{\vbox{\baselineskip12pt
%\advance\hsize by -1truein\noindent\footnotefont{\bf Fig.~\the\figno:} #2}}
%\bigskip\endinsert
\global\advance\figno by1} \fi

%%%%%%% References %%%%%%%

\def\det{{\rm det}}

%% MACROS

\def\IL{\relax{\rm I\kern-.18em L}}
\def\IH{\relax{\rm I\kern-.18em H}}
\def\IR{\relax{\rm I\kern-.18em R}}
\def\IC{\relax\hbox{$\inbar\kern-.3em{\rm C}$}}
\def\IZ{\relax\ifmmode\mathchoice
{\hbox{\cmss Z\kern-.4em Z}}{\hbox{\cmss Z\kern-.4em Z}}
{\lower.9pt\hbox{\cmsss Z\kern-.4em Z}} {\lower1.2pt\hbox{\cmsss
Z\kern-.4em Z}}\else{\cmss Z\kern-.4em Z}\fi}

\def\CN {{\cal N}}

\def\CL {{\cal L}}

\def\CO {{\cal O}}
\def\CZ {{\cal Z}}

%% MORE MACROS

\def\CN {{\cal N}}

\def\CO {{\cal O}}

\def\CZ {{\cal Z }}

\def\det{{\rm det}}
\def\Tr{{\rm Tr}}

\font\manual=manfnt \def\dbend{\lower3.5pt\hbox{\manual\char127}}

\def\IZ{\relax\ifmmode\mathchoice
{\hbox{\cmss Z\kern-.4em Z}}{\hbox{\cmss Z\kern-.4em Z}}
{\lower.9pt\hbox{\cmsss Z\kern-.4em Z}} {\lower1.2pt\hbox{\cmsss
Z\kern-.4em Z}}\else{\cmss Z\kern-.4em Z}\fi}

\def\lfm#1{\medskip\noindent\item{#1}}

\def\bar{\overline}

\def\rt2{\sqrt{2}}
\def\irt2{{1\over\sqrt{2}}}

%  \slashchar puts a slash through a character to represent contraction
%  with Dirac matrices. Use \not instead for negation of relations, and use
%  \hbar for hbar.
\def\slashchar#1{\setbox0=\hbox{$#1$}           % set a box for #1
   \dimen0=\wd0                                 % and get its size
   \setbox1=\hbox{/} \dimen1=\wd1               % get size of /
   \ifdim\dimen0>\dimen1                        % #1 is bigger
      \rlap{\hbox to \dimen0{\hfil/\hfil}}      % so center / in box
      #1                                        % and print #1
   \else                                        % / is bigger
      \rlap{\hbox to \dimen1{\hfil$#1$\hfil}}   % so center #1
      /                                         % and print /
   \fi}

%\TeschnerYF
\lref\TeschnerYF{ J.~Teschner, ``On the Liouville three point
function,'' Phys.\ Lett.\ B {\bf 363}, 65 (1995)
[arXiv:hep-th/9507109].
%%CITATION = HEP-TH 9507109;%%
}

%\FateevIK
\lref\FateevIK{ V.~Fateev, A.~B.~Zamolodchikov and
A.~B.~Zamolodchikov, ``Boundary Liouville field theory. I:
Boundary state and boundary  two-point function,''
arXiv:hep-th/0001012.
%%CITATION = HEP-TH 0001012;%%
}

%\DouglasUP
\lref\DouglasUP{ M.~R.~Douglas, I.~R.~Klebanov, D.~Kutasov,
J.~Maldacena, E.~Martinec and N.~Seiberg, ``A new hat for the $c =
1$ matrix model,'' arXiv:hep-th/0307195.
%%CITATION = HEP-TH 0307195;%%
}

%\MooreMG
\lref\MooreMG{ G.~W.~Moore, ``Geometry Of The String Equations,''
Commun.\ Math.\ Phys.\  {\bf 133}, 261 (1990).
%%CITATION = CMPHA,133,261;%%
}

%\MooreCN
\lref\MooreCN{ G.~W.~Moore, ``Matrix Models Of 2-D Gravity And
Isomonodromic Deformation,'' Prog.\ Theor.\ Phys.\ Suppl.\  {\bf
102}, 255 (1990).
%%CITATION = PTPSA,102,255;%%
}

%\KlebanovWG
\lref\KlebanovWG{ I.~R.~Klebanov, J.~Maldacena and N.~Seiberg,
``Unitary and complex matrix models as 1-d type 0 strings,''
Commun.\ Math.\ Phys.\  {\bf 252}, 275 (2004)
[arXiv:hep-th/0309168].
%%CITATION = HEP-TH 0309168;%%
}
%\JohnsonHY
\lref\JohnsonHY{ C.~V.~Johnson, ``Non-perturbative string
equations for type 0A,'' arXiv:hep-th/0311129.
%%CITATION = HEP-TH 0311129;%%
}
%\SeibergNM
\lref\SeibergNM{ N.~Seiberg and D.~Shih, ``Branes, rings and
matrix models in minimal (super)string theory,'' JHEP {\bf 0402},
021 (2004) [arXiv:hep-th/0312170].
%%CITATION = HEP-TH 0312170;%%
}

%\BleherYS
\lref\BleherYS{ P.~Bleher and A.~Its, ``Double scaling limit in
the random matrix model: the Riemann-Hilbert approach,''
arXiv:math-ph/0201003.
%%CITATION = MATH-PH 0201003;%%
}

%\MaldacenaSN
\lref\MaldacenaSN{ J.~Maldacena, G.~W.~Moore, N.~Seiberg and
D.~Shih, ``Exact vs. semiclassical target space of the minimal
string,'' arXiv:hep-th/0408039.
%%CITATION = HEP-TH 0408039;%%
}

%\CrnkovicWD
\lref\CrnkovicWD{ C.~Crnkovic, M.~R.~Douglas and G.~W.~Moore,
``Loop equations and the topological phase of multi-cut matrix
models,'' Int.\ J.\ Mod.\ Phys.\ A {\bf 7}, 7693 (1992)
[arXiv:hep-th/9108014].
%%CITATION = HEP-TH 9108014;%%
}

%\ItsKG
\lref\ItsKG{ A.~R.~Its, A.~S.~Fokas and A.~A.~Kapaev, ``On the
asymptotic analysis of the Painleve equations via the isomonodromy
method,'' INS-233
%\href{http://www.slac.stanford.edu/spires/find/hep/www?r=ins-233}{SPIRES entry}
}

%\SenNF
\lref\SenNF{ A.~Sen, ``Tachyon dynamics in open string theory,''
arXiv:hep-th/0410103.
%%CITATION = HEP-TH 0410103;%%
}

%\HollowoodXQ
\lref\HollowoodXQ{ T.~J.~Hollowood, L.~Miramontes, A.~Pasquinucci
and C.~Nappi, ``Hermitian versus anti-Hermitian one matrix models
and their hierarchies,'' Nucl.\ Phys.\ B {\bf 373}, 247 (1992)
[arXiv:hep-th/9109046].
%%CITATION = HEP-TH 9109046;%%
}

%\DalleyBR
\lref\DalleyBR{ S.~Dalley, C.~V.~Johnson, T.~R.~Morris and
A.~Watterstam, ``Unitary matrix models and 2-D quantum gravity,''
Mod.\ Phys.\ Lett.\ A {\bf 7}, 2753 (1992) [arXiv:hep-th/9206060].
%%CITATION = HEP-TH 9206060;%%
}

%\HoravaJY
\lref\HoravaJY{ P.~Horava, ``Type IIA D-branes, K-theory, and
matrix theory,'' Adv.\ Theor.\ Math.\ Phys.\  {\bf 2}, 1373 (1999)
[arXiv:hep-th/9812135].
%%CITATION = HEP-TH 9812135;%%
}

%\KutasovFG
\lref\KutasovFG{ D.~Kutasov, K.~Okuyama, J.~w.~Park, N.~Seiberg
and D.~Shih, ``Annulus amplitudes and ZZ branes in minimal string
theory,'' JHEP {\bf 0408}, 026 (2004) [arXiv:hep-th/0406030].
%%CITATION = HEP-TH 0406030;%%
}

%\MaldacenaXJ
\lref\MaldacenaXJ{ J.~M.~Maldacena, G.~W.~Moore and N.~Seiberg,
``D-brane instantons and K-theory charges,'' JHEP {\bf 0111}, 062
(2001) [arXiv:hep-th/0108100].
%%CITATION = HEP-TH 0108100;%%
}

%\GrossHE
\lref\GrossHE{ D.~J.~Gross and E.~Witten, ``Possible Third Order
Phase Transition In The Large N Lattice Gauge Theory,'' Phys.\
Rev.\ D {\bf 21}, 446 (1980).
%%CITATION = PHRVA,D21,446;%%
}

%\JohnsonUT
\lref\JohnsonUT{ C.~V.~Johnson, ``Tachyon condensation,
open-closed duality, resolvents, and minimal bosonic and type 0
strings,'' arXiv:hep-th/0408049.
%%CITATION = HEP-TH 0408049;%%
}

%\GaiottoNZ
\lref\GaiottoNZ{ D.~Gaiotto, L.~Rastelli and T.~Takayanagi,
``Minimal superstrings and loop gas models,''
arXiv:hep-th/0410121.
%%CITATION = HEP-TH 0410121;%%
}

%\KapustinPK
\lref\KapustinPK{ A.~Kapustin, ``A remark on worldsheet fermions
and double-scaled matrix models,'' arXiv:hep-th/0410268.
%%CITATION = HEP-TH 0410268;%%
}

%\KawaiPJ
\lref\KawaiPJ{ H.~Kawai, T.~Kuroki and Y.~Matsuo, ``Universality
of nonperturbative effect in type 0 string theory,''
arXiv:hep-th/0412004.
%%CITATION = HEP-TH 0412004;%%
}

%\FukudaBV
\lref\FukudaBV{ T.~Fukuda and K.~Hosomichi, ``Super Liouville
theory with boundary,'' Nucl.\ Phys.\ B {\bf 635}, 215 (2002)
[arXiv:hep-th/0202032].
%%CITATION = HEP-TH 0202032;%%
}
%\AhnEV
\lref\AhnEV{ C.~Ahn, C.~Rim and M.~Stanishkov, ``Exact one-point
function of N=1 super-Liouville theory with boundary,'' Nucl.\
Phys.\ B {\bf 636}, 497 (2002) [arXiv:hep-th/0202043].
%%CITATION = HEP-TH 0202043;%%
}

\lref\WakimotoKac{ V.~G.~Kac and M.~Wakimoto, ``Exceptional
hierarchies of soliton equations," Proc. Sympos. Pure Math. {\bf
49}, 191 (1989).}

\lref\MorrisBW{ T.~R.~Morris, ``2-D Quantum Gravity, Multicritical
Matter And Complex Matrices,'' FERMILAB-PUB-90-136-T
%\href{http://www.slac.stanford.edu/spires/find/hep/www?r=fermilab-pub-90-136-t}{SPIRES entry}
}

%\DiFrancescoRU
\lref\DiFrancescoRU{ P.~Di Francesco, ``Rectangular Matrix Models
and Combinatorics of Colored Graphs,'' Nucl.\ Phys.\ B {\bf 648},
461 (2003) [arXiv:cond-mat/0208037].
%%CITATION = COND-MAT 0208037;%%
}

\lref\PeriwalGF{ V.~Periwal and D.~Shevitz, ``Unitary Matrix
Models As Exactly Solvable String Theories,'' Phys.\ Rev.\ Lett.\
{\bf 64}, 1326 (1990);
%%CITATION = PRLTA,64,1326;%%
V.~Periwal and D.~Shevitz, ``Exactly Solvable Unitary Matrix
Models: Multicritical Potentials And Correlations,'' Nucl.\ Phys.\
B {\bf 344}, 731 (1990).
%%CITATION = NUPHA,B344,731;%%
}

\lref\CrnkovicMS{ C.~Crnkovic, M.~R.~Douglas and G.~W.~Moore,
``Physical Solutions For Unitary Matrix Models,'' Nucl.\ Phys.\ B
{\bf 360}, 507 (1991).
%%CITATION = NUPHA,B360,507;%%
}

%\NappiBI
\lref\NappiBI{ C.~R.~Nappi, ``Painleve-II And Odd Polynomials,''
Mod.\ Phys.\ Lett.\ A {\bf 5}, 2773 (1990).
%%CITATION = MPLAE,A5,2773;%%
}

%\DalleyQG
\lref\DalleyQG{ S.~Dalley, C.~V.~Johnson and T.~Morris,
``Multicritical complex matrix models and nonperturbative 2-D
quantum gravity,'' Nucl.\ Phys.\ B {\bf 368}, 625 (1992).
%%CITATION = NUPHA,B368,625;%%
}

%\LafranceWY
\lref\LafranceWY{ R.~Lafrance and R.~C.~Myers, ``Flows For
Rectangular Matrix Models,'' Mod.\ Phys.\ Lett.\ A {\bf 9}, 101
(1994) [arXiv:hep-th/9308113].
%%CITATION = HEP-TH 9308113;%%
}
%\BrowerMN
\lref\BrowerMN{ R.~C.~Brower, N.~Deo, S.~Jain and C.~I.~Tan,
``Symmetry breaking in the double well Hermitian matrix models,''
Nucl.\ Phys.\ B {\bf 405}, 166 (1993) [arXiv:hep-th/9212127].
%%CITATION = HEP-TH 9212127;%%
}

\lref\FateevIK{V.~Fateev, A.~B.~Zamolodchikov and
A.~B.~Zamolodchikov, ``Boundary Liouville field theory. I:
Boundary state and boundary two-point function,''
arXiv:hep-th/0001012.
%%CITATION = HEP-TH 0001012;%%
}

\lref\ZamolodchikovAH{ A.~B.~Zamolodchikov and
A.~B.~Zamolodchikov, ``Liouville field theory on a pseudosphere,''
arXiv:hep-th/0101152.
%%CITATION = HEP-TH 0101152;%%
}

\lref\Itsbook{ A.~R.~Its and V.~Y.~Novokshenov, ``The
Isomonodromic Deformation Method in the Theory of Painlev\'e
Equations," Springer-Verlag (1986).}

%\FlaschkaWJ
\lref\FlaschkaWJ{ H.~Flaschka and A.~C.~Newell, ``Monodromy And
Spectrum Preserving Deformations. 1,'' Commun.\ Math.\ Phys.\ {\bf
76}, 65 (1980).
%%CITATION = CMPHA,76,65;%%
}

%\MorozovHH
\lref\MorozovHH{ A.~Morozov, ``Integrability And Matrix Models,''
Phys.\ Usp.\  {\bf 37}, 1 (1994) [arXiv:hep-th/9303139].
%%CITATION = HEP-TH 9303139;%%
}

\lref\Jimbo{ M.~Jimbo, T.~Miwa and K.~Ueno, ``Monodromy preserving
deformation of linear orderinary differential equations with
rational coefficients, I," Physica D {\bf 2}, 306 (1981)
}

\lref\ItsNovok{ A.~R.~Its and V.~Yu.~Novokshenov, ``Effective
sufficient conditions for the solvability of the inverse problem
of monodromy theory for systems of linear ordinary differential
equations," Funct.\ Anal.\ Appl.\ {\bf 22}, 190 (1988)
   }

\lref\ItsKapaev{ A.~R.~Its, A.~A.~Kapaev, ``The Irreducibility of
the Second Painleve Equation and the Isomonodromy Method," in
   {\it Toward the exact WKB analysis of differential equations,
   linear or non-linear}, C.~Howls, T.~Kawai and Y.~Takei, eds.,
   Kyoto University Press (2000).
}

%\draftmode

\newbox\tmpbox\setbox\tmpbox\hbox{\abstractfont }
\Title{\vbox{\baselineskip12pt \hbox{PUPT-2148}}}
{\vbox{\centerline{Flux Vacua and Branes}\smallskip \centerline{of
the Minimal Superstring}}}
\smallskip
\centerline{Nathan Seiberg$^1$ and David Shih$^2$}
\smallskip
\bigskip
\centerline{$^1${\it School of Natural Sciences, Institute for
Advanced Study, Princeton, NJ 08540 USA}}
\medskip
\centerline{$^2${\it Department of Physics, Princeton University,
Princeton, NJ 08544 USA}}
\bigskip
\vskip 1cm

\noindent We analyze exactly the simplest minimal superstring
theory, using its dual matrix model. Its target space is one
dimensional (the Liouville direction), and the background fields
include a linear dilaton, a possible tachyon condensate, and RR
flux. The theory has both charged and neutral branes, and these
exhibit new and surprising phenomena. The smooth moduli space of
charged branes has different weakly coupled boundaries in which
the branes have different RR charges. This new duality between
branes of different charges shows that the semiclassical notion of
localized charge is not precise in the quantum theory, and that
the charges of these branes can fluctuate. Correspondingly, the RR
flux in some parts of target space can also fluctuate -- only the
net flux at infinity is fixed. We substantiate our physical
picture with a detailed semiclassical analysis of the exact
answers. Along the way, we uncover new subtleties in
super-Liouville theory.

\vskip 1cm

\Date{December, 2004} \vfil\eject

\newsec{Introduction}

Minimal superstring theories, or $(p,q)$ superminimal CFT coupled
to $\CN=1$ super-Liouville theory, are interesting toy models of
string theory. Dual to certain large $N$ matrix models
\refs{\GrossHE\PeriwalGF\NappiBI\CrnkovicMS\CrnkovicWD\HollowoodXQ
\BrowerMN\MorrisBW\DalleyQG\DalleyBR\LafranceWY-\DiFrancescoRU},
they have a precise and tractable nonperturbative definition. (For
recent work on minimal superstrings, see
\refs{\KlebanovWG\JohnsonHY\SeibergNM\JohnsonUT\GaiottoNZ\KapustinPK-\KawaiPJ}.)
Along with their bosonic cousins, they exhibit many important
stringy phenomena, such as D-branes, holography and open/closed
duality. In addition, the minimal superstring allows us to study
interesting phenomena associated with RR flux and RR charge. The
study of such flux vacua is interesting, among other reasons,
because of their importance in the landscape of more generic
string vacua.

In this paper, we will focus on the simplest minimal superstring
theory, corresponding to $(p,q)=(2,4)$. We will use the exact
matrix model description to study nonperturbative aspects of the
theory in RR flux backgrounds. Specializing to $(p,q)=(2,4)$
simplifies the analysis considerably. In particular, it allows us
to use interchangeably the 0A and 0B formulations of the theory,
because in this case the two are equivalent \KlebanovWG. We will
see that some observables are easier to analyze in one formulation
or the other.

One of the nice features of minimal string theory is that it
allows us to study in great detail the interplay between various
dual descriptions -- worldsheet, target space, and matrix model --
of the theory. These descriptions all have their strengths and
weakness. For instance, calculations are easiest to carry out
using the exact matrix model description, but their physical
interpretation is often obscure. Conversely, the physics is
clearest in the target space description, as it resembles that of
more generic models, but calculations are often impossible using
this description. Finally, the worldsheet formalism allows us to
analyze the semiclassical physics to all orders in $\alpha'$, but
the higher order quantum corrections are not easy to calculate,
and the exact answers are impossible to find.

Since our ultimate goal is to extract target space physics from
the matrix model, let us describe in some detail the system from
the target space perspective, following closely \KlebanovWG. The
target space consists of a single dimension $\phi$, described by
the Liouville field on the worldsheet. The linear dilaton
background for $\phi$ means that the string coupling depends on
$\phi$ as
 \eqn\stringc{g_s=e^{Q \phi \over 2} = e^{3 \phi \over 2\sqrt{2}}
 }
where $Q={3\over\sqrt{2}}$ is the linear dilaton slope. Thus $g_s$
is zero at $\phi\to -\infty$ and grows exponentially as $\phi\to
+\infty$. We will refer to these as the weak and strong coupling
regions, respectively.

The target space fields consist of the closed-string tachyon
$T(\phi)$ and the RR scalar $C(\phi)$. In the vacuum we have
 \eqn\Tvev{
T(\phi)=  \mu e^{b \phi }= \mu e^{\phi \over \sqrt 2}
 }
where $\mu$ is referred to as the worldsheet bulk cosmological
constant, and $b={1\over\sqrt{2}}$ is the worldsheet Liouville
coupling constant. In order to examine the value of $C(\phi)$ in
the vacuum, we consider the leading term in its Lagrangian
\refs{\DouglasUP,\KlebanovWG}
 \eqn\LagrangianC{
  \CL_C = {1\over 2\sqrt{2}} e^{-2T}(\partial_\phi C)^2
 }
This Lagrangian has the following symmetries: it is invariant
under charge conjugation $C\to -C$ and under constant shifts of
$C$. The latter gives rise to a conserved current
 \eqn\fluxdef{q= e^{-2T} \partial_\phi C}
We identify it as RR flux, which originates from RR charge $q$.

The solutions to the equations of motion of \LagrangianC\ are
 \eqn\classicalC{
 C(\phi) = C_0 + q \int^\phi e^{2T(\phi')} d\phi'}
where $q$ is the background flux \fluxdef.  The norm of the small
fluctuations of $C$ is derived from \LagrangianC, $||\delta
C||^2=\int d\phi e^{-2T}(\delta C)^2$.  It determines the nature
of the various possible fluctuations. For example, for positive
$\mu$ the mode $C_0$ in \classicalC\ is (delta function)
normalizable.  This fluctuating mode allows us to explore the
total charge of the system.  This is to be contrasted with the
flux term $q \int^\phi e^{2T(\phi')} d\phi'$ which is not
normalizable as $\phi \to +\infty$. Such deformations were
identified in \KutasovFG\ as arising from branes which are
localized at infinity \ZamolodchikovAH. In the worldsheet
description these are charged ZZ branes \refs{\FukudaBV,\AhnEV}.
For negative $\mu$, things are slightly different. Here the flux
term is normalizable in the strong coupling region, showing that
the flux does not originate from a charged brane at infinity. The
fact that this term is linearly divergent as $\phi \to -\infty$
shows that it is controlled by boundary conditions in the
asymptotic weak coupling end, and in the worldsheet description of
the theory it is described by a vertex operator (the RR ground
state). By contrast, the solution $C_0$ is not normalizable at the
strong coupling end, and therefore it does not correspond to an
allowed vertex operator. It cannot fluctuate and it labels a
superselection sector. However, because of the shift symmetry of
$C$, the physical answers are independent of $C_0$. This is
similar to the situation with spontaneous symmetry breaking. The
novelty here, which happens because of the special coupling
\LagrangianC, is that we have spontaneous symmetry breaking in one
dimension.

For positive $\mu$, the flux term in \classicalC\ arises from
charged branes at infinity, so we cannot explore its contribution
to the classical action without knowing more about these branes.
We will do this below. However, for negative $\mu$, we can simply
substitute the classical solution \classicalC\ into \LagrangianC\
to find \KlebanovWG
 \eqn\actionC{ S_C = {1\over2\sqrt{2}}q^2 \int_{-\infty}^{\infty}
 d\phi\,
 \exp(2\mu e^{\phi\over\sqrt{2}}) ={1\over 2}q^2\log\left(\Lambda/|\mu|\right) }
where $\Lambda$ is a cutoff on $\phi$ in the weak coupling region.
Later we will compare this with the matrix model results.

Besides analyzing bulk observables in flux backgrounds, we will
also study D-branes and their interplay with the flux. Minimal
superstring theory has both charged and neutral stable branes,
which in the Liouville literature are called FZZT branes
\refs{\FateevIK,\TeschnerMD}.  Their super-Liouville version was
explored in \refs{\FukudaBV,\AhnEV}. These branes are labelled by
the value of the worldsheet boundary cosmological constant
$\mu_B$. In target space, they have an open string tachyon whose
value
 \eqn\Topenvevintro{ T_{open}(\phi)=\mu_B e^{b\phi\over 2} = \mu_B e^{\phi\over 2\sqrt
 2} }
interpolates between $T_{open}=0$ as $\phi\to -\infty$ and
$T_{open}=\pm\infty$ as $\phi\to +\infty$. The minisuperspace
wavefunction $\Psi_{open}(\phi) \sim e^{-( T_{open})^2}$
\DouglasUP\ means that we can think of these branes as
semiextended, stretching from the weak coupling region to a point
$\phi=\phi_0\sim -2\sqrt{2}\log |\mu_B|$, where they dissolve
away. We will argue below that semiclassically, the charged branes
carry a half-unit of RR charge,
 \eqn\chargexrel{
  q_b = {1\over2}{\rm sign}(\mu_B)
  }
localized around $\phi=\phi_0$. Together with flux $q$ in the
strong coupling region, this gives rise to flux
 \eqn\qweakintro{ q_{weak}=q+q_b}
in the weak coupling region. The wavefunction of the brane is
shown together with its localized charge in figure 1.

\medskip \ifig\figI{The minisuperspace wavefunction for the charged FZZT brane,
along with its localized charge. The brane comes in from the weak
coupling region $\phi\to -\infty$ and dissolves at an intermediate
point $\phi=\phi_0\sim -\log |\mu_B|$. By varying $\mu_B$, we can
bring the tip of the brane into the weak or strong coupling
region.
}
    {\epsfxsize=0.6\hsize\epsfbox{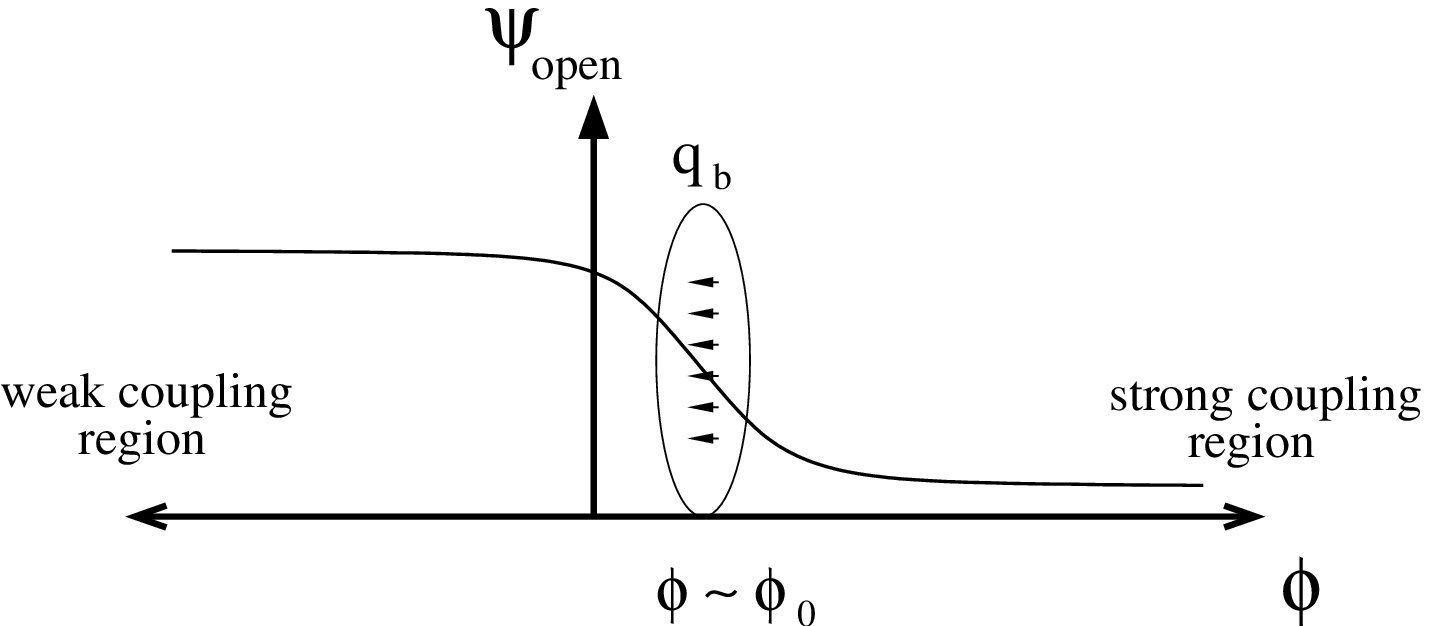}}

It may seem strange that we can change the charge of the brane by
changing the sign of $\mu_B$. According to \qweakintro, this would
also appear to change the value of $q_{weak}$. But for $\mu_B
\approx 0$, the charge of the brane is localized in the
strong-coupling region, and it is hard to see how varying $\mu_B$
by a small amount could affect the value of the flux in the
weak-coupling region. This suggests that as $\mu_B$ is varied
across zero, the value of the flux in the strong coupling region
changes by one unit, in such a way that
$q_{weak}=q+q_b=(q+2q_b)-q_b$ is preserved. But this means that
neither the flux in the strong coupling region nor the charge of
the brane is well-defined! At large $|\mu_B|$, these quantities
freeze to their semiclassical values, but in general they can
fluctuate. Only $q_{weak}$, the flux in the weak-coupling region,
is unambiguous.

Below we will discuss these phenomena in detail. We will also see
how this qualitative target space picture is realized very
quantitatively in the dual matrix model. In the matrix model
description, we start with a closed-string background labelled by
$q$, and there are initially two charged branes $B_{s}(x)$,
$s=\pm{1\over2}$, which lead to flux
\eqn\qweakintroii{
q_{weak}=q+s
}
in the weak-coupling region. The parameter $x$ which is natural in
the matrix model is related to $\mu_B$ by
\eqn\xdefintro{
x=i\mu_B
}
So real $\mu_B$ corresponds to $x$ along the imaginary axis.
Notice that the branes in the matrix model are essentially
labelled by the flux in the weak coupling region \qweakintroii,
which is well-defined. This is to be expected, since the matrix
model provides an exact description of the system. Now, using
differential equations derived from the matrix model, it is
possible to analyze the exact partition functions
 \eqn\psidefintro{ \psi_{\pm}(x) = {\langle B_{\pm {1/2}}(x)
 \rangle_{\mu,q}\over \CZ(\mu,q)}
 }
of the charged branes. Here $\langle B_{\pm
1/2}(x)\rangle_{\mu,q}$ is the unnormalized FZZT partition
function (defined by the matrix integral), while $\CZ(\mu,q)$ is
the bulk partition function. We will show that these partition
functions satisfy the nontrivial identity \eqn\psiidentityintro{
\langle B_{s}(x)\rangle_{\mu,q} = \langle
B_{-s}(x)\rangle_{\mu,q+2s} } This identity substantiates the
target space picture described above in a very precise way. It
says that there is really only one brane for each $x$, and that we
can think of this brane as either having charge $s$ in background
flux $q$, or charge $-s$ in background flux $q+2s$. The only
unambiguous quantity is the net flux $q_{weak}=q+s$ at weak
coupling.

These charged FZZT branes are analogous to the unstable branes of
the critical string (for a review and a list of references see
\SenNF).  In addition to these branes our system also has neutral
extended branes.  These are analogous to the unstable
brane-anti-brane states of the critical string.  The open string
tachyon on these neutral branes is complex and its phase is a
gauge degree of freedom.  Unlike their critical string
counterparts, our charged and neutral branes are stable because
the open string tachyon on them is actually massive.

The outline of the paper is as follows. In section 2 we will
review and extend the analysis of the closed-string sector of our
system. Semiclassically, it has two phases, distinguished by the
sign of the cosmological constant $\mu$, and separated by a
third-order phase transition \GrossHE. The exact theory, however,
does not exhibit a transition; rather, the interpolation from
positive to negative $\mu$ is smooth
\refs{\PeriwalGF\NappiBI\CrnkovicMS\CrnkovicWD\HollowoodXQ-\BrowerMN,\KlebanovWG}.
This has an interesting consequence in backgrounds with nonzero
flux $q$: as $\mu$ is varied, there is a smooth transition from a
phase where the flux comes from charged branes at infinity, to a
phase where the flux comes from closed strings. Finally, we derive
in section 2 various nontrivial identities for the bulk free
energy that are necessary for the ensuing analysis of the
D-branes.

In section 3 we review some of the facts of the worldsheet theory,
focusing on the FZZT branes in super-Liouville theory.  In
particular, we will present some new terms which have to be added
to the boundary states of some of the branes. These additions are
motivated by our results in later sections.

In sections 4 and 5, we study the charged and neutral D-branes of
the system. Using differential equations derived from the matrix
model, we will analyze their partition functions in various
limits. As in \MaldacenaSN, the exact answers are entire functions
of $x$, but due to Stokes' phenomenon the semiclassical limit
exhibits monodromy. We will see that in the superstring, Stokes'
phenomenon is much more intricate and has additional physical
consequences not present in the bosonic string. This additional
structure arises largely from the possibility of having RR charge
and flux in the minimal superstring. Thus, the study of the exact
answers will lead to a much better understanding of background RR
fluxes and their interplay with the various kinds of D-branes of
the theory.

Finally, a few appendices are devoted to various technical details
and relations to different points of view.

\newsec{The closed-string sector}

\subsec{The equations}

In this section, we will review some results on the closed-string
sector of our model, following closely the presentation in
\KlebanovWG.\foot{As mentioned in the introduction, the 0A and 0B
GSO projections are equivalent for $(p,q)=(2,4)$, so we will not
distinguish between them. In \KlebanovWG, 0A and 0B were
identified with a change in sign of the cosmological constant
$\mu$. We will stick to the 0A sign conventions of \KlebanovWG.}
The observables of the closed-string sector depend on the values
of the cosmological constant $\mu$ and the RR flux $q$. Using the
duality with the matrix model (which we review in appendix A), one
can derive differential equations -- called the ``string
equations" -- for the bulk partition function $\CZ(\mu,q)$. These
equations are most conveniently formulated in terms of a function
$r=r(\mu,q)$, which is related to the free energy
$F(\mu,q)=-\log\CZ(\mu,q)$ via
 \eqn\partf{{1\over2}r^2 =
 F''
 }
In terms of $r$, the string equations take the form:
 \eqn\plII{\eqalign{
 &r'' - \mu r - r^3 +
 r\beta'^2=0 \cr
 &r^2\beta' ={q}}}
where $'$ is shorthand for $\partial_\mu$ and $\beta=\beta(\mu,q)$
is an auxiliary function. As we will see in a moment, $\beta$ is
useful because it makes manifest certain symmetries of the system.
Of course, it can be easily eliminated, leading to a differential
equation for $r$ alone:
 \eqn\plIIq{r'' - \mu r - r^3
 + {q^2 \over r^3}=0}
We recognize this as a modified version of the Painlev\'e II
equation. Note that the equations \plII\ can be interpreted as the
equations of motion derived from the ``Lagrangian''
 \eqn\actionII{{1\over 2}(r')^2 - {1\over 2}r^2 \beta'^2 + \
 {1\over 2}\mu r^2 + {1\over 4}r^4  + q \beta'}
where we view $\beta'$ as an independent variable. Integrating out
$\beta'$ results in a Lagrangian that describes a particle with
``time" $\mu$ and ``position" $r(\mu)$, moving in a
one-dimensional potential
 \eqn\rpot{ V(r)=-\left({1\over 4}r^4+{1\over2}\mu r^2
 +{q^2\over2r^2}\right) }
From the form of the potential, it is not hard to see that for
every $q$ there exists a solution to the equations of motion which
is everywhere real and positive, and which interpolates smoothly
between $r=+\infty$ at $\mu=-\infty$ and $r=0$ at $\mu=+\infty$.
These are the physical solutions that we will focus on henceforth.

The equations of motion \plII\ and the action \actionII\ are
invariant under the charge conjugation symmetry: $\beta \to
-\beta$, $q\to -q$. It follows that
 \eqn\qmqsym{ r(\mu,q) = r(\mu,-q) }
(Actually, the symmetry could have related two different
solutions, but this is not the case for the solution satisfying
our boundary conditions.) In addition, \plII\ and \actionII\ are
also invariant under shifts of $\beta$ by a constant. The
parameter $q$, which appears in \actionII\ as a ``topological
term,'' can be interpreted as the conserved charge associated with
this symmetry. Below we will limit ourselves to $q\in \Bbb Z $ and
will make a few comments about more general values.

Now let us explore some of the properties of these equations. For
this, it is convenient to define
 \eqn\Zpmd{Z_{\pm}= r e^{\mp \beta}= r e^{\mp \int {q\over r^2}
 d\mu}}
Notice that the charge conjugation symmetry \qmqsym\ implies
 \eqn\chgconjZ{ Z_+(\mu,q)=Z_-(\mu,-q) }
The solutions to the string equation satisfy a nontrivial identity
involving $Z_\pm$ \refs{\HollowoodXQ,\WakimotoKac}
 \eqn\clstridentity{
 r(\mu,q\pm1)^2 = r(\mu,q)^2-
 2(\log Z_\pm(\mu,q))'' }
It is amusing to view \clstridentity\ as a discrete version of KdV
flow, with the discrete parameter $q$ playing the role of the
coupling constant. To prove \clstridentity, simply check that the
square root of the RHS solves the string equation \plIIq\ with
$q\to q\pm 1$. Of course, this does not guarantee that the
solution $r(\mu,q\pm 1)$ generated in this way is physical, in the
sense described below \rpot. In the next subsection, we will see
that $q\in \Bbb Z$ is the only consistent set of $q$ for which
this is always true.

Using \partf, we can integrate \clstridentity\ twice and
exponentiate to obtain\foot{There are possible $q$-dependent
integration constants here. We show in appendix B that these can
be consistently set to zero.}
 \eqn\ZFidentity{Z_\pm(\mu,q)=e^{F(\mu,q)-F(\mu,q\pm1)}=
 {\CZ(\mu,q\pm 1) \over \CZ(\mu,q)}}
Thus $Z_\pm(\mu,q)$ can be thought of as the expectation value of
an operator which changes $q \to q \pm 1$. Finally, let us mention
a trivial consequence \ZFidentity:
\eqn\clstridentityiv{ Z_+(\mu,q-1)Z_-(\mu,q) = 1 } This expression
will be useful below.

\subsec{Semiclassical limit}

Since $g_s \sim |\mu|^{-3/2}$ according to \stringc\ and \Tvev,
the semiclassical $g_s\to 0$ limit corresponds to the $|\mu| \to
\infty $ limit. For $q=0$ the leading terms in the semiclassical
expansion are
 \eqn\claplII{\eqalign{
  &r(\mu,q=0) =\cases{\sqrt{-{\mu} }(1 +
 \CO(\mu^{-3})) & $\mu<0$ \cr
 \sqrt{2}Ai(\mu)(1 + \CO(e^{-{4 \mu^{3/2}\over 3}}))  &
 $\mu>0$}\cr
  &\beta(\mu,q=0)=0 \cr
  &Z_{\pm}(\mu,q=0) = r(\mu,q=0)
 }}
We have set $\beta=0$ using the freedom to shift $\beta$ by a
constant, in order to preserve the charge conjugation symmetry
\chgconjZ\ at $q=0$.

For nonzero $q$ the situation is more interesting.  For negative
$\mu$ (the two cut phase of the 0B theory) the solution of the
string equations \plII\ is
 \eqn\solqab{\eqalign{
 &r(\mu,q) = \sqrt{-{\mu }}(1+ {1-4q^2\over 8\mu^3}+
 \CO(\mu^{-6} ))\cr
 &\beta(\mu,q)=-q \log(-\mu)  + \CO(\mu^{-3} )\cr
 &Z_\pm(\mu,q) = r e^{\mp \beta}=(-\mu)^{\pm q+1/2}
 (1+ \CO(\mu^{-3} ))}}
Here we have again used the freedom to shift $\beta$ by a constant
in order to satisfy \clstridentityiv.

For positive $\mu$ (the one cut phase of the 0B theory) the limit
$q\to 0$ is not smooth. We will find it convenient throughout to
parametrize this discontinuity in terms of a function
\eqn\epsqdef{
 \epsilon(q)=\cases{-1 & $q<0$\cr
 0 & $q=0$\cr
 +1 & $q>0$}
}
Then for nonzero $q$ we find
 \eqn\solqa{\eqalign{
 &r(\mu,q) = \sqrt{|q| \over \sqrt\mu}(1 -{|q| \over 4 \mu^{3/2}}
 + \CO(\mu^{-3}) )\cr
 &\beta(\mu,q)=\epsilon(q){2 \mu^{3/2} \over 3} + {q \over 2} \log(
 \mu)+ \log B(q)+ \CO(\mu^{-3/2})\cr
 &Z_\pm(\mu,q) = r e^{\mp \beta}=B(q)^{\mp1}\sqrt{|q| }(
 \mu)^{-{1\over 4}\mp {q \over 2}}e^{\mp\epsilon(q){2 \mu^{3/2} \over 3}}(1+
 \CO(\mu^{-3/2}))\cr
 }}
Note that in order to find the leading order contributions to
$Z_\pm$ we need to expand $\beta(\mu,q)$ (and hence $r(\mu,q)$) to
the next to leading order. In \solqa\ we have included a possible
integration constant $B(q)$ in $\beta$. This cannot be set to
zero, since we have already used this freedom at the $\mu\to
-\infty$ end \solqab. Instead, we can use the fact that
$Z_+(\mu,q-1)Z_-(\mu,q)$ is independent of $\mu$ to determine
$B(q)$ -- since we have set $Z_+(\mu,q-1)Z_-(\mu,q)=1$ at $\mu\to
-\infty$, the same must be true at $\mu\to +\infty$.

Using these asymptotic expansions we can, as promised, examine the
compatibility between the identity \clstridentity\ and the
physical requirements on $r(\mu,q)$ described below \rpot.
Consider what happens for $q\in(-1,0)$. If \clstridentity\ holds,
then for large positive $\mu$, using \solqa\ gives
 $r(\mu,q+1)^2 \sim -{q+1\over \mu^{1/2}} < 0$.
But this violates the physical requirement on $r$ that it be
everywhere real and positive.  So if we want to include noninteger
values of $q$, we are forced to discard the identity
\clstridentity. Conversely, if we take
 \eqn\assumequant{ q\in \Bbb Z }
then \clstridentity\ is always satisfied. We will see below that
\clstridentity\ has desirable, physical consequences. Therefore,
we will assume $q\in \Bbb Z$ throughout the paper.

It is also interesting to calculate from \solqab\ and \solqa\ the
leading order terms in the perturbative expansion of the bulk free
energy. Here we find\foot{A similar discussion of the integration
constants applies here.  Since $F(\mu,q)$ is smooth for real
$\mu$, we fix the integration constants (to zero) at $\mu \to
-\infty$ using \ZFidentity\ and then the integration constant
$A(q)$ is determined also for $\mu \to +\infty$.}
 \eqn\Fleadingcalc{ F(\mu,q) = \cases{ -{\mu^3\over
12}+{1\over8}\log(-\mu)-{1\over2}q^2\log(-\mu)+\CO(\mu^{-3}) &
$\mu<0$\cr
 {2\over3}|q|\mu^{3/2}+{1\over4}q^2\log\mu + A(q)+ \CO(\mu^{-3/2}) &
 $\mu>0$}
 }
As a consistency check, notice that the leading-order
$q$-dependent term for $\mu<0$ agrees precisely with the
calculation \actionC\ based on the target space action. From
\Fleadingcalc, we can learn about the physical origin of the flux
$q$ in the two phases. In general, we expect the genus expansion
 \eqn\Fgenus{ F = \sum_{h,b} F_{h,b}\, |\mu|^{3(1-h)}\left({|q|
 \over |\mu|^{3\over 2}}\right)^b }
with $F_{h,b}$ independent of $\mu$ and $q$.  Here $h$ is the
number of handles in the worldsheet and $b$ is the number of
boundaries. According to \Fleadingcalc, the leading contribution
for $\mu>0$ is from $h=0$ and $b=1$, i.e.\ the disk. The factor of
$|q|$ indicates that there are $q$ charged branes on which the
boundary of the worldsheet can end. The higher order corrections
are also consistent with this picture. We identify these branes
with the charged ZZ branes localized at strong coupling that exist
in this phase. Meanwhile, for $\mu<0$ all the terms in the
perturbative expansion \Fgenus\ have even $b$.  We interpret this
to mean that they come from surfaces with no boundaries but with
$2b$ insertions of an RR vertex operator with coefficient $q$. For
instance, the leading order $q^2$ term comes from two insertions
of the RR vertex operator on the sphere.

Note that for $\mu$ positive, $q$ arises from charged instantons.
However, it does not fluctuate, because it affects the flux at
infinity. This is to be contrasted with the bosonic string, where
the analogous contributions must be summed, because there the
instantons do not carry charge. From \Fleadingcalc, we also see
that the instantons have action ${2\over3}\mu^{3/2}$ in this
phase. For $q=0$, the leading order nonperturbative effects (see
\claplII) are $\CO(e^{-{4 \mu^{3/2} \over 3}})$; thus, these are
instanton-anti-instanton effects. As in the bosonic string, these
instanton-anti-instanton effects should be summed over because
they do not have charge.

As we cross from positive to negative $\mu$, there is a transition
in the strong-coupling region between the flux $q$ being generated
by charged branes or simply being there without branes as sources.
Since the exact answer is a smooth function of $\mu$, this
transition must be smooth as well.

In the rest of the paper, we will study the extended branes of the
minimal superstring -- known as FZZT branes on the worldsheet --
and their interaction with RR flux backgrounds described in this
section. We will see that these branes provide us with further
interesting examples of smooth, nonperturbative transitions
between seemingly different flux vacua.

\newsec{Worldsheet Description of FZZT Branes}

Before going on to describe the exact matrix model analysis of the
FZZT branes, let us first review how they are described on the
worldsheet. The worldsheet description will provide nontrivial
checks of the exact answer, as well as shed light on its physical
interpretation.

In the worldsheet description, D-branes are conveniently described
using boundary states. The boundary states for the FZZT branes
were first written down for super-Liouville theory in
\refs{\FukudaBV,\AhnEV}, and they were later adapted to the
minimal superstring in \refs{\DouglasUP,\KlebanovWG,\SeibergNM}.
They depend on the sign of $\mu$, and they are labelled by a
parameter $\eta=\pm 1$ which determines the boundary condition on
the worldsheet supercharge, $Q=i\eta \bar Q$. For $\mu>0$, they
are\foot{Our conventions here differ slightly with those of
\SeibergNM: $\sigma_{here}={\pi b\over2}\sigma_{there}$,
$P_{here}={1\over b}P_{there}$, and $\mu_{here}=-2\mu_{there}$. }
  \eqn\bsfzztmup{\eqalign{
    &|\sigma,\eta=-1\rangle_{\pm} =
     \int_{0}^{\infty}dP\, \left({\mu\over2}\right)^{iP} \Big( \cos(2
    P\sigma)|P,\eta=-1\rangle\rangle_{NS}
    \pm \cos(2P\sigma)|P,\eta=-1\rangle\rangle_{R}\Big)\cr
  &|\sigma,\eta=+1\rangle =\sqrt{2}\int_{0}^{\infty}dP\,
    \left({\mu\over2}\right)^{iP} \cos(2P\sigma)
    |P,\eta=+1\rangle\rangle_{NS}
 }}
Here $|P,\eta\rangle\rangle_{NS,R}$ are the NS and R Ishibashi
states, apart from an overall $P$ dependent normalization that we
will ignore. The RR Ishibashi states in the branes with $\eta=-1$
show that they are charged, while branes with $\eta=+1$ are
neutral.

Notice that the relative normalization between the $\eta=+1$ and
$\eta=-1$ boundary states differs by a factor of $\sqrt{2}$. This
factor of $\sqrt 2$ originates from the ratio of the disk of the
spin field Cardy state and the identity and fermion Cardy states
in the Ising model.  It is also common in the similar setup of
non-BPS branes in the critical string \refs{\HoravaJY,\SenNF}. We
will confirm it below in the exact analysis.

The boundary states are most natural in terms of $\sigma$, but the
physical parameter of interest is the boundary cosmological
constant $\mu_B$. The relation between the two
\refs{\FukudaBV,\AhnEV} depends on $\eta$ (and, as we will see,
the sign of $\mu$):
 \eqn\etamub{
 \mu_B = -i x = \cases{ \sqrt{\mu\over 2}\cosh \sigma & $\eta=-1$ \cr
                  \sqrt{\mu\over2} \sinh \sigma & $\eta=+1$\cr}
 }
In \SeibergNM, this relation was interpreted as the boundary
analogue of the (super)B\"acklund transformation, with $\sigma$
identified as the Dirichlet boundary condition on the B\"acklund
field. Note that here we have also introduced $x=i\mu_B$, the
parameter most natural in the matrix model description.

Using the boundary states, we can compute the genus zero one-point
functions of physical vertex operators simply by substituting the
appropriate value of $P$ (for details, see \SeibergNM). In
particular, the one-point function of the cosmological constant
operator (given by $P=-{i\over2}$) is simply
\eqn\ccopf{
\left\langle \int d^2z\,e^{{\phi(z)\over\sqrt{2}}}\right\rangle =
\sqrt{\mu\over2}\cosh(\sigma)
 }
for both signs of $\eta$. Integrating once with respect to $\mu$
holding $x$ fixed yields the FZZT disk amplitude
\refs{\KlebanovWG,\SeibergNM}: \eqn\bsdiskmup{
 D(x) = \cases{ -i\left({4\over3}x^3+\mu x\right) & $\eta=-1$\cr
                -{4\over3}\sqrt{2}\left({\mu\over2}-x^2\right)^{3/2}
                & $\eta=+1$\cr} }
Notice that, strictly speaking, the integral with respect to $\mu$
does not determine the $x^3$ term for $\eta=-1$. We will see how
this is fixed below, when we study the disk for $\mu<0$.

The disk amplitudes can be written in the form $D(x)=\int^x y(x')
dx'$, with
 \eqn\diskymup{
 \qquad y(x)^2=\cases{ -(4x^2+\mu)^2 &\quad $\eta=-1$\cr
                                            -32\,x^2(x^2-{\mu\over2})
                                            &\quad $\eta=+1$\cr
                                            }
 }
The algebraic curve $y^2=y(x)^2$ defines a Riemann surface
associated with each FZZT brane. This Riemann surface is a double
cover of the complex $x$ plane. For $\eta=-1$, this surface breaks
into two subsurfaces where $y=\pm y(x)$. The subsurfaces are
connected by two singularities at imaginary $x=\pm
{\sqrt{-\mu}\over2}$. Meanwhile, for $\eta=+1$ the surface is
irreducible, has branch points at $x=\pm\sqrt{\mu\over2}$ (it is
convenient to take the cuts to lie along the real $x$ axis), and
has a singularity at $x=0$. It was shown in \SeibergNM\ how these
surfaces provide a simple geometric interpretation for minimal
superstring theory.  In particular, the two subsurfaces with $\eta
=-1$ represent the branes with the two different charges
\SeibergNM.  Below we will see how this classical picture is
modified in the exact theory.

Now let us turn to the FZZT boundary states for $\mu<0$. These are
given by
 \eqn\bsfzztmun{\eqalign{
 &|\sigma,\eta=-1\rangle_{\pm} =\cr &\quad \int_{0}^{\infty}dP\, \left(-{\mu\over2}\right)^{iP} \Big(
    \cos(2P\sigma)| P,\eta=-1\rangle\rangle_{NS}
   \mp i\sin (2P\sigma)|
    P,\eta=-1\rangle\rangle_{R}\Big)\pm {1\over2}V_R|0\rangle\cr
 &|\sigma,\eta=+1\rangle =\sqrt{2}
\int_{0}^{\infty}dP\,\left(-{\mu\over2}\right)^{iP}
    \cos(2P \sigma)|P,\eta=+1\rangle\rangle_{NS}
    \cr
 }}
where now
 \eqn\etamubii{
 \mu_B = -i x = \cases{ \sqrt{-{\mu\over2}}\sinh \sigma & $\eta=-1$ \cr
                 \sqrt{-{\mu\over2}}\cosh \sigma & $\eta=+1$\cr}
 }
Here $V_R$ is the RR vertex operator which creates one unit of
flux. Its $\phi$ dependence in the $(-1/2,-3/2)$ picture is $\phi
e^{Q\phi/2}$, and in the $(-1/2,-1/2)$ picture it is simply
$e^{Q\phi/2}$. The extra term $\pm {1\over2}V_R|0\rangle$ was
missed in the literature. The exact analysis in the next section
will show that it is necessary. Note that if not for this extra
term, the state $|\sigma,\eta=-1\rangle_+$ would be identical to
the state $|-\sigma,\eta=-1\rangle_-$. This was the basis of the
claim in the literature that these two $\eta=-1$ FZZT branes are
identical in this phase. The new term shows that this claim is in
fact false. Instead, the worldsheet theory has two oppositely
charged $\eta=-1$ branes, for both signs of $\mu$.

As before, we can derive the disk amplitude starting from the
one-point function of the cosmological constant operators. This
gives
\eqn\bsdiskmun{
 D(x) = \cases{ -{4\over3}i\left(x^2+{\mu\over2}\right)^{3/2} & $\eta=-1$\cr
                 -i\sqrt{2}\left({4\over3}x^3-\mu x\right) &
 $\eta=+1$\cr}
 }
Here we have determined the $x^3$ term and the overall
normalization of the $\eta=+1$ disk amplitude as follows. When
$|x|$ is large, the brane is far in the weak coupling region, so
the disk amplitude should be insensitive to the sign of $\mu$, as
this only affects the physics in the strong-coupling region. In
the $\mu>0$ phase \bsdiskmup, the $\eta=+1$ disk becomes $D(x)\to
-i\sqrt{2}({4\over3}x^3-\mu x)$ at large $|x|$. This fixes the
$x^3$ term and the overall normalization of the $\eta=+1$ disk in
the $\mu<0$ phase, \bsdiskmun. A similar argument also fixes the
$x^3$ term and the overall normalization of the $\eta=-1$ disk
amplitude in \bsdiskmup, as promised.

We can again write the disk amplitude as $D(x)=\int^x y(x') dx'$
with
\eqn\diskymun{
 \qquad y(x)^2=\cases{ -16x^2(x^2+{\mu\over 2}) &\quad $\eta=-1$\cr
                                            -2(4x^2-\mu)^2
                                            &\quad $\eta=+1$\cr
                                            }
 }
Again, $y^2=y(x)^2$ defines Riemann surfaces for each $\eta$.
Comparing with \diskymup, it is obvious that the $\mu<0$,
$\eta=\pm1$ surfaces are essentially the same as the $\mu>0$,
$\eta=\mp1$ surfaces. However, we will see later that the
different $\eta$ surfaces are distinguished by their response to
nonzero $q$.

Let us also point out that the new term $\pm
{1\over2}V_R|0\rangle$ in \bsfzztmun\ does not affect the disk
amplitude.  Therefore it does not affect the curve $y(x)$.
However, since it distinguishes between
$|\sigma,\eta=-1\rangle_{+}$ and $|-\sigma,\eta=-1\rangle_{-}$,
these two branes are not related by analytic continuation in
$\sigma$, and they take values in two different but isomorphic
surfaces. Below, we will return to this point.

In the next section, we will study the exact FZZT partition
functions, which reduce to $e^{D(x)+\dots}$ in the semiclassical
limit. Since $D(x)$ takes values on a Riemann surface, the
semiclassical limit exhibits monodromy. However, we will see that
the exact answers are entire in the complex $x$ plane. As in
\MaldacenaSN, Stokes' phenomenon replaces the semiclassical
Riemann surface with a single copy of the $x$ plane.

\newsec{The Charged Branes}

\subsec{The differential equations}

In this section, our goal is to study the exact partition
functions \eqn\psidefBZ{ \psi_\pm = \psi_\pm(x,\mu,q)={\langle
B_{\pm 1/2}(x)\rangle_{\mu,q}\over\CZ(\mu,q)} } for the charged
branes $B_{\pm 1/2}$ of our system. These are related to the
$\eta=-1$ FZZT branes in the worldsheet description reviewed
above. We will see, however, that the relationship is subtle in
many ways.

The charged branes are most natural in the 0B language. Through
the identification of the FZZT branes with the determinant
operator of the 0B matrix model (see appendix A.1), one can derive
the following differential equations for $\psi_\pm$:
 \eqn\flco{\eqalign{
 &\left(\partial_\mu - A_\mu\right)\Psi=
 \left(\partial_\mu - \pmatrix{-i x &  {Z_+\over\sqrt{2}} \cr
 {Z_-\over\sqrt{2}}& i x }\right)\Psi = 0 \cr
  &\left(\partial_x - A_x\right)\Psi =\left(\partial_x-
  \pmatrix{-i(4 x^2 +r^2 + \mu )
 & 2\sqrt{2}xZ_++\sqrt{2}i Z_+'\cr 2\sqrt{2}xZ_--\sqrt{2}i Z_-'
 & i(4 x^2 +r^2 + \mu )  }\right) \Psi=0 \cr
 }}
where $\Psi = \left(\matrix{\psi_+ \cr \psi_-}\right)$. These
equations were studied extensively for $q=0$ in \Itsbook, and the
generalization to nonzero $q$ is straightforward. The details are
left to appendix A.1.  Here let us just mention three consistency
checks. First, notice that in the normalization where $Z_\pm$ have
charges $\pm 1$, the functions $\psi_\pm$ have charges $\pm
{1/2}$. This is consistent with the target space interpretation of
the branes, and it justifies our notation $B_{\pm 1/2}$ above.
Second, it is important that the string equation \plIIq\ implies
the flatness condition \eqn\fieldstrength{
 F_{x\mu}= \partial_x A_\mu - \partial_\mu A_x
 +\left[A_\mu,A_x \right]=0
 }
This in turn guarantees the compatibility of the two equations in
\flco. Third, the fact that $A_\mu$ and $A_x$ are regular for all
(finite) $x\in \Bbb C$ and $\mu\in \Bbb R$ means that
$\Psi(x,\mu,q)$ is an entire function of $x$ and is smooth for
$\mu$ real. This agrees with the expectation that the branes
should be single-valued with respect to the coupling constants.

In order to solve the differential equations \flco, we must
specify boundary conditions for $\Psi$. The two linearly
independent solutions to \flco\ are distinguished by their
behavior at large $|x|$, with one exponentially increasing and the
other exponentially decreasing as $x\to +i\infty$. On the other
hand, the physical solution must satisfy
\eqn\bczb{
\lim_{x\to +i\infty}\Psi(x,\mu,q)= \lim_{x\to
-i\infty}\Psi(x,\mu,q)=0
}
This follows from the fact that the matrix model potential goes to
$+\infty$ at $x\to \pm i\infty$ (see the discussion in appendix
A.1). For $q=0$, the existence of a regular solution to \flco\
satisfying \bczb\ was proven in \BleherYS. We can extend the
existence proof to all $q\in \Bbb Z$ by using the following
nontrivial identity
 \eqn\psiidentity{ \psi_-(x,\mu,q) =
 Z_-(\mu,q)\psi_+(x,\mu,q- 1) }
which relates solutions with different $q$. Note that we can also
write \psiidentity\ as
$\psi_+(x;\mu,q)=Z_+(\mu,q)\psi_-(x;\mu,q+1)$ using
\clstridentityiv. Starting from the physical solution at $q=0$,
these identities generate solutions at $q=\pm 1$ with the correct
asymptotics. This proves the existence of the physical solution
for $q=\pm1$, and continuing in this way we prove existence for
all $q\in \Bbb Z$. In the process, we also determine the overall
$q$-dependent normalization of $\psi_\pm$ which is left unfixed by
\flco.

The identity \psiidentity\ has been studied before in the
mathematical literature (see e.g.\ \refs{\ItsNovok,\Jimbo}) and is
sometimes referred to as a Schlesinger or B\"acklund
transformation. Below we will understand the physical
interpretation of \psiidentity\ in terms of D-branes and fluxes.
Since the proof of \psiidentity\ is rather technical, we reserve
it for appendix B.

Another property of the solutions follows from the charge
conjugation symmetry of the theory. As mentioned above, the bulk
theory is invariant under $q\to -q$. Since
$Z_+(\mu,q)=Z_-(\mu,-q)$, charge conjugation is also a symmetry of
the differential equations \flco, provided we interchange $\psi_+$
with $\psi_-$ and send $x\to -x$. However, a symmetry of the
equations is not necessarily a symmetry of the solutions. Instead,
charge conjugation could map one solution into another.
Fortunately, this does not happen here. Charge conjugation must
map the physical solution at $q$ into the physical solution at
$-q$, since the unphysical solution is exponentially {\it
increasing} at large imaginary $x$. Thus we conclude that
 \eqn\psiC{\psi_+(-x,\mu,q)=\psi_-(x,\mu,-q)}

Let us explore some of the consequences of these identities.
First, notice that we can use \psiidentity\ to rewrite the
differential equations as equations for $\psi_+$ alone:
\eqn\flcorew{\eqalign{
 &\partial_\mu\tilde\Psi= \pmatrix{-i x &  {1\over\sqrt{2}}r^2 \cr
 {1\over\sqrt{2}}& i x - \partial_\mu \log Z_-}\tilde\Psi\cr
  & \partial_x \tilde\Psi =
  \pmatrix{-i(4 x^2 +r^2 + \mu )
 & r^2\left(2\sqrt{2}x  +\sqrt{2}i\partial_\mu \log Z_+\right)\cr
 2\sqrt{2}x -\sqrt{2}i \partial_\mu\log Z_-
 & i(4 x^2 +r^2 + \mu )  }
 \tilde\Psi \cr
 }}
where $\tilde\Psi = \left(\matrix{\psi_+(x,\mu,q)\cr
\psi_+(x,\mu,q-1)}\right)$ and all the quantities in matrices are
evaluated at $q$. Eq.\ \flcorew\ makes manifest the fact that
$\psi_+$ is the only independent function in this analysis.

A second, more interesting consequence of these identities follows
from using \psidefBZ\ and the identity \ZFidentity\ to rewrite
\psiidentity\ and its charge conjugate as
 \eqn\samesys{ \langle B_{s}\rangle_{\mu,q} = \langle B_{-s}\rangle_{\mu,q+2s}
 }
with $s=\pm 1/2$. Evidently, a system with closed string
background labelled by $q$ and a brane $B_{s}$ is the same as a
system with a closed string background labelled by $q+2s$ with a
brane $B_{-s}$! In other words, the distinct configurations of the
system are labelled only by the value of $q+s$, and not separately
by $q$ and $s$. In the next subsection, we will understand this
fact in the target space language, where $q+s$ is identified as
the value of the RR flux in the asymptotic weak coupling region.

\subsec{Target space interpretation}

As discussed in the introduction, the charged FZZT branes have an
open string tachyon whose value is
 \eqn\opentachyon{ T_{open}(\phi)=\mu_B e^{{\phi\over2\sqrt{2}}}
 = -ix\, e^{\phi\over2\sqrt{2}}}
The minisuperspace wavefunction $\Psi_{open}(\phi)$ shows that the
brane comes in from $\phi=-\infty$ and dissolves at
 \eqn\phizero{ \phi = \phi_0 \approx -{2\sqrt{2}} \log (|x|) }
(Recall figure 1.) We will now argue that its charge is localized
near $\phi_0$.

The key observation is that the charged FZZT branes are in many
ways analogous to the unstable branes \refs{\HoravaJY,\SenNF} of
critical string theory. In that context $T_{open}=0$ is an
unstable situation which can decay to one of two stable minima
$T_{open}=\pm T_0$. The theory on these branes has a coupling to
the bulk RR scalar $C$,
 \eqn\nonbpscoupling{ \delta \CL =  C(\phi)\,\partial_\phi
 G(T_{open}(\phi)) }
where $G(T_{open})$ is some function of the open string tachyon
field. The charge conjugation symmetry $C \to - C$, $T_{open}\to
-T_{open}$ shows that the function $G$ is odd and hence $G(0)=0$.
Furthermore, $G(\pm T_0)=\pm {1\over 2}$, so that a kink
interpolating between the two minima carries one unit of RR
charge.

Now consider our system. Here the point $T_{open}=0$ is locally
stable, and the open string configuration \opentachyon\
interpolates between $T_{open}=0$ as $\phi \to -\infty$ and
$T_{open}= \pm \infty$ as $\phi \to + \infty$. Hence the FZZT
brane corresponds to ``half a kink" and carries charge
 \eqn\qbdef{ q_b =  {1\over 2} \epsilon(\mu_B)= {1 \over 2}
 \epsilon({\rm Im}\,\,x) }
which is localized around $\phi \approx \phi_0$.

Notice that the sign of the charge depends on whether $x$ is in
the UHP or the LHP rather than on the value of $s$ in $B_{s}$.
This immediately leads to two questions:

 \lfm{1.} How can this be consistent with the fact that
the expectation values $\psi_\pm$ are smooth functions of $x$?
That is, how can the charge on the brane change as in \qbdef?
 \lfm{2.} How is the target space charge $q_b$ related to the
 matrix model charge $s$?

\medskip

In order to address these questions, let us examine the target
space situation more closely.  Consider a background with flux
$q$, and introduce into this background a brane with charge $q_b$;
i.e.\ the value of $x$ is such that $q_b = {1 \over 2}
\epsilon({\rm Im}\,\,x)$.  We start with a brane with large $|{\rm
Im}\,\,x|$ so that it does not penetrate much into the strong
coupling region $\phi \to +\infty$. Then in a first approximation,
this brane does not affect the situation in the strong coupling
region; in particular the flux in the strong coupling region
remains the original flux,
 \eqn\qstrongi{ q_{strong}= q }
Around the point $\phi_0$ we have charge $q_b$, and
correspondingly the flux in the weak coupling region $\phi \to
-\infty$ is \eqn\qweak{ q_{weak}=q +q_b } Now, as we make $x$
small and pass through $x=0$ the charge of the brane should change
to $-q_b$. Since nothing happens at weak coupling, we expect that
the flux in the weak coupling region remains $q_{weak}=q +q_b$.
This means that the flux in the strong coupling region should
change to \eqn\qstrongii{ q_{strong}= q+2 q_b } The branes and the
transition between them are depicted in figure 2.

\medskip \ifig\figII{The target space configuration corresponding
to the charged branes. Recall that $\mu_B=-ix$. Semiclassically,
the brane charge $q_b={1\over2}\epsilon(\mu_B)$ changes sign and
the strong coupling flux $q_{strong}$ jumps by one unit as we pass
from large positive to large negative $\mu_B$. However, the two
flux configurations are smoothly connected in the exact theory.}
    {\epsfxsize=0.8\hsize\epsfbox{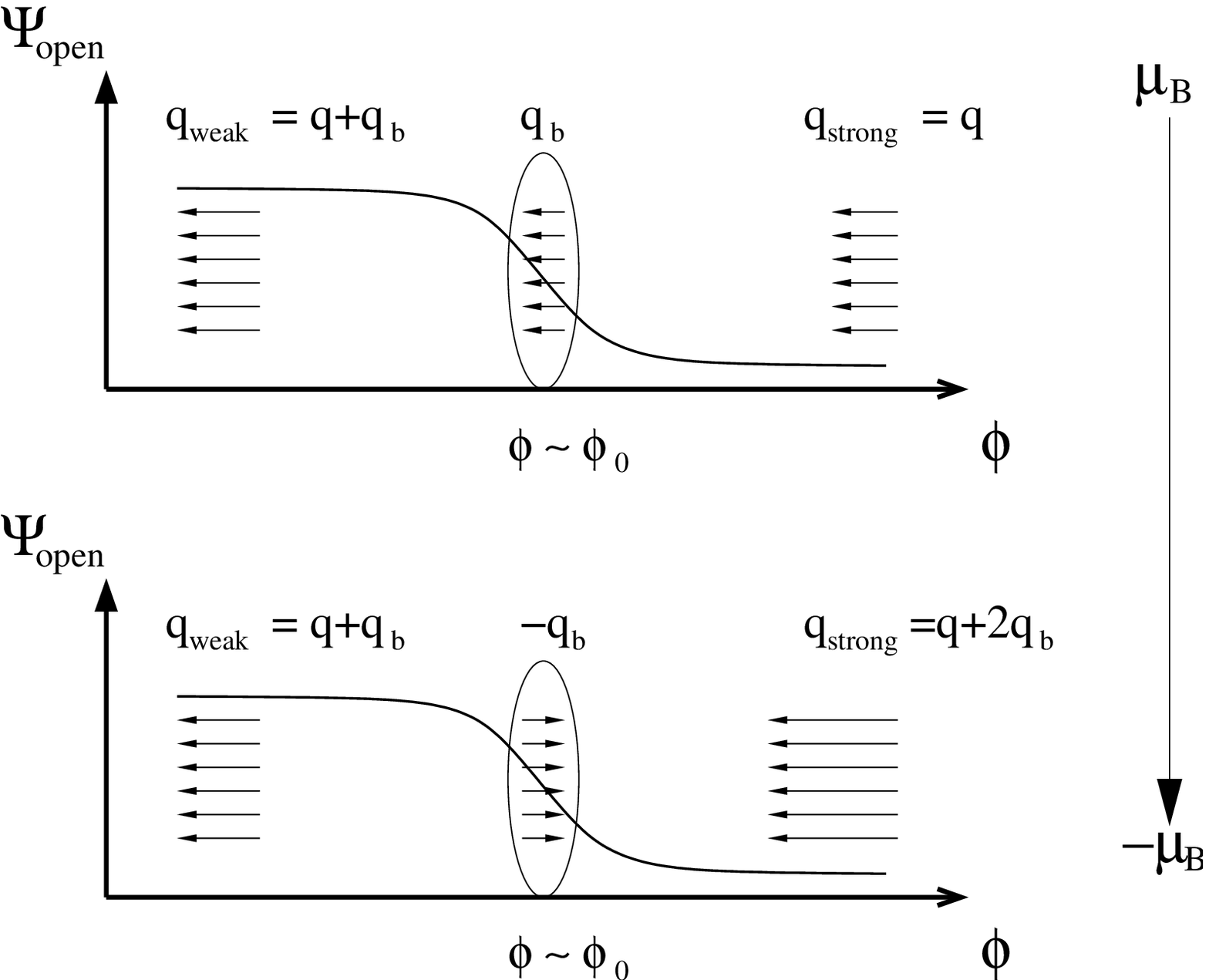}}

Returning now to the first question above, evidently the surprise
is that this transition must happen smoothly. Clearly, this means
that the charge $q_b$ and the flux $q_{strong}$ cannot jump
abruptly. Instead, they must be fluctuating quantities which are
in general ill defined. For large $|x|$ they freeze at the values
we discuss above, but for small $|x|$ they are not well defined.
This is to be contrasted with the flux at the weak coupling end
$q_{weak}$, which does not change. It is specified by the boundary
conditions on the system.

We interpret this result to mean that the effective volume of the
strong coupling region is finite. Hence the flux there and the
charge on the brane can fluctuate. On the other hand, the weak
coupling region has infinite volume and hence the flux there is
fixed.  As $|x|$ becomes larger, the volume of the strong coupling
region grows, and the fluctuations are reduced. Therefore $q_b$
and $q_{strong}$ freeze in this limit.

The physical picture presented here motivates the addition of the
term $\pm {1\over 2} V_R|0\rangle$ in the $q=0$ boundary state
\bsfzztmun.\foot{We thank J.~Maldacena for a discussion on this
point.}  Consider a closed string configuration with $q=0$ and
place in it the brane created by the naive boundary state without
the extra term. If this brane has large $x$, it is far in the
weak-coupling region. Then we do not expect it to significantly
affect the physics in the strong-coupling region, i.e.\
$q_{strong}=q=0$. But a simple calculation shows that this cannot
be true if we use the naive boundary state.  The flux in the
weak-coupling region is measured by the one point function of the
RR ground state vertex operator on the disk.  Since the
coefficient $\sin(2P\sigma)$ of the RR part of the boundary state
vanishes at zero momentum $P$, this one point function vanishes.
Hence $q_{weak}=0$ and correspondingly, $q_{strong} = \mp {1\over
2}$. This contradicts our argument above that $q_{strong} = 0$.
Moreover, even if we allowed such flux in the strong coupling
region, its value $\mp {1 \over 2}$ would conflict with our
assumption of flux quantization. These inconsistencies are cured
by the added term $\pm {1\over2}V_R|0\rangle$, which adds flux
$\pm {1\over2}$ everywhere, restoring $q_{strong}=0$ and leading
to $q_{weak}=\pm {1\over2}$.

As a check of this picture of the target space dynamics, consider
again the brane identifications \samesys. The identifications were
presented in section 4.1 as a consequence of the differential
equations \flco, but now we see that they fit nicely with our
understanding of target space. According to \samesys, the same
configuration can be thought of as a brane with charge $q_b=s$ and
background flux $q_{strong}=q$, or as a brane with charge $q_b=-s$
and background flux $q_{strong}=q+2s$. The brane charge and the
strong coupling flux are not meaningful quantities in general. The
only invariant is the net value of the flux at the weak coupling
end, $q_{weak}=q+s$. This agrees precisely with our target space
understanding.

\medskip \ifig\figIII{The target space configuration corresponding
to $B_{\pm 1/2}$ for $x$ in the LHP and the UHP. The figure makes
clear the fact that the label $s$ in $B_s$ distinguishes between
different values of the weak coupling flux $q_{weak}=q+ s$, and
not the semiclassical brane charge $q_b$. }
    {\epsfxsize=0.9\hsize\epsfbox{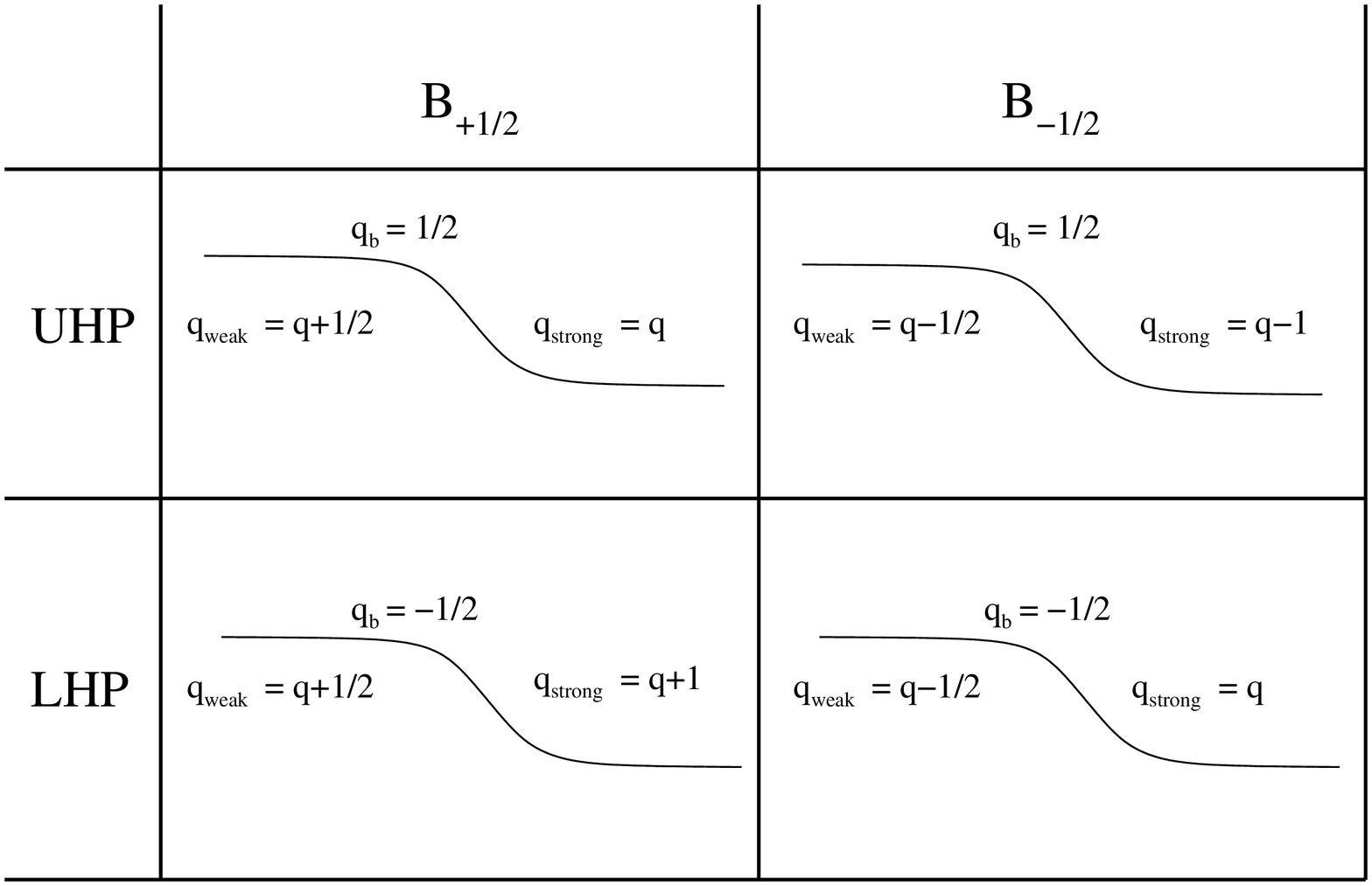}}

The brane identifications \samesys\ also allow us to answer the
second question above, namely the relation between the target
space charge $q_b$ and the matrix model charge $s$. Clearly, the
branes $B_{+1/2}$ and $B_{-1/2}$ only have their naive
semiclassical charge assignments $q_b=1/2$ and $q_b=-1/2$ for $x$
in the UHP and LHP, respectively. The identifications \samesys\
ensure that analytically continuing these branes to the other half
plane changes the sign of $q_b$ and changes $q_{strong}$ by one
unit of flux, in such a way that $q_{weak}$ is preserved. These
four configurations are shown in figure 3.

To summarize, the various situations are labelled by $q_{weak} \in
{\Bbb Z}+ {1\over 2}$.  For large $|x|$ the charge of the brane
and the flux in the strong coupling region are meaningful.  They
are $q_b={1\over 2} \epsilon({\rm Im}\,\, x)$, and
$q_{strong}=q_{weak}- q_b$. But they are not meaningful for finite
$|x|$, and hence a smooth transition between these two situations
is possible. It is worth mentioning that this identification of
branes is reminiscent of the identification of brane charges in
the context of K-theory which was discussed in \MaldacenaXJ.

\subsec{Large $|x|$ asymptotics -- weak coupling limit}

We will devote the next few subsections to solving the
differential equations \flco\ in various asymptotic limits. This
will lead to some checks and a better understanding of the
physical picture we have just described.

The simplest limit is $|x| \to \infty$ with fixed $\mu$.  Since
$|x|$ is large, the branes are dissolved in the weak coupling
region, and therefore we expect the various fluxes and charges to
be frozen. Furthermore, since $\mu$ is taken here to be of order
one, the closed string background at the point the brane is
dissolved is simply a linear dilaton background, and
correspondingly the worldsheet theory is free.

We generalize the results of \refs{\Itsbook,\BleherYS} to nonzero
$q$
 \eqn\sliuhp{\pmatrix{\psi_+ \cr \psi_-}  \sim \cases{
  (-2\sqrt{2}ix)^{q} e^{-i({4 \over 3 } x^3 + \mu x)}
  \pmatrix{1 +
  \CO(1/x)\cr -{Z_-(\mu,q) \over 2\sqrt{2}ix} + \CO(1/x^2)} &
 $0<\arg(x) <\pi$\cr
 \qquad & \cr \qquad & \cr
  (2\sqrt{2}ix)^{-q} e^{i({4 \over 3 } x^3 + \mu x)}
  \pmatrix{{Z_+(\mu,q) \over 2\sqrt{2}ix} + \CO(1/x^2) \cr
  1+ \CO(1/x)}&
 $-\pi <\arg(x) <0$\cr}}
In the wedges $|\arg(x)| < \pi/3$ and $|\arg(x) -\pi| <\pi/3$ we
must take the sum of the result in the UHP and LHP. Note that,
although we have used the identity \psiidentity\ to fix the
overall $q$-dependent normalization of $\psi_\pm$, it is still
nontrivial that the $x$ and $\mu$ dependence of \sliuhp\  respect
\psiidentity\ and \psiC.

Let us explore
the expressions in more detail.  For $x$ in the UHP we have
$|\psi_+| \gg |\psi_-|$, while for $x$ in the LHP the opposite is true.
We take this to mean that in the UHP (LHP) the
naive classical picture is correct for the brane $B_{+1/2}$ ($B_{-1/2}$).
This confirms the target space picture discussed above.
We identify the power of $x$ in the
asymptotic expansion \sliuhp\ as $2q_b q_{strong}$. We will see
below how this dependence can be understood from a worldsheet
description.

\subsec{Semiclassical limit}

Further insight into these phenomena arises by studying the
semiclassical $g_s\to 0$ limit of $\psi_\pm$. As noted above
\claplII, this limit corresponds to $|\mu|\to \infty$. In
addition, we must take $|x| \to \infty$  holding
 \eqn\txdef{\tilde x = {x \over\sqrt{\mu}} = {\rm fixed}
 }
This way, the branes still dissolve in the weak coupling region,
although now the tachyon background is nonzero there.
Correspondingly, the worldsheet theory is interacting. This will
allow us to explore world sheet phenomena, but with frozen
background charges and fluxes.  We will see that the naive
semiclassical picture does not always lead to the correct answer.
Instead, the correct semiclassical pictures come with different
values of the flux, depending on whether we are in the UHP or the
LHP. After taking this subtlety into account, we find complete
agreement with worldsheet expectations.

Let us focus on the upper half plane for simplicity. The answers
in the lower half plane can be easily obtained using \psiC. In the
semiclassical limit, the answers depend on the sign of $\mu$. For
positive $\mu$ (the one cut phase of the 0B theory) we have
 \eqn\sliuhpqu{\eqalign{
 &\pmatrix{\psi_+ \cr \psi_-}  \sim
 e^{-i({4 \over 3 } x^3 + \mu x)}\pmatrix{ Y(x,q)^{q}
 (1 + \CO(\mu^{- 3/2}))\cr
  Z_-(\mu,q )Y(x,q-1)^{q-1}  \left(1 + \CO(\mu^{- 3/2})\right) }\cr
}}
where
\eqn\Ydef{
  Y(x,q)=-i2\sqrt{2}\left(x+{i\over 2}
  \epsilon(q)\sqrt\mu \right)
 }
and $\epsilon(q)$ was defined in \epsqdef. In \sliuhpqu, and later
when we discuss the situation with negative $\mu$, the meaning of
the correction terms like $\CO(\mu^{-3/2})$ is that they are
multiplied by a function of $\tilde x=x/\sqrt{\mu}$.

It is easy to see that in the limit $|x| \gg |\mu| \gg 1$ these
expressions are consistent with the large $|x|$ expression
\sliuhp. Comparing with \sliuhp, we see that the only effect of
large $\mu$ is to shift $x$ in the prefactor by
$x_{ZZ}=-{i\over2}\epsilon(q)\sqrt{\mu} $. This has the following
consequence: the semiclassical approximation \sliuhpqu\ has a
branch point at $x=x_{ZZ}$ if $q$ is not an integer. This is
another example that the solutions are better behaved for integer
$q$.

The physical (worldsheet) interpretation of the semiclassical
limit \sliuhpqu\ is illuminating. Keep in mind that in the $\mu>0$
phase, we can have charged ZZ branes. The semiclassical limit of
$\psi_+$ in the UHP is consistent with a disk amplitude
 \eqn\diskmup{ D(x)=-i\left({4 \over 3 } x^3 + \mu x\right) }
This disk amplitude agrees with the results in section 3 and in
\refs{\KlebanovWG,\SeibergNM}. The asymptotics of $\psi_+$ also
imply that the annulus between the FZZT brane and itself is zero,
 \eqn\annmupff{ Z_{annulus}(x,x)=0 }
while the log of the prefactor
 \eqn\annmupfq{ \tilde Z_{annulus}(x,q) = q\log Y(x,q) }
can be interpreted as due to an annulus diagram with one end on
the FZZT brane and the other on $|q|$ ZZ branes.  More precisely,
note that $Y(x,q)$ can be written as
 \eqn\Yrew{\eqalign{
 &Y = -i2\sqrt{2} (x-x_{ZZ}(q)) \cr
 &x_{ZZ}(q)=-{i\over 2} \epsilon(q) \sqrt\mu}}
where the value of $x_{ZZ} $ is the position of the ZZ branes on
the semiclassical Riemann surface \refs{\KlebanovWG,\SeibergNM}.
Notice that all of the $|q|$ background ZZ branes are located at
just one of the two singularities of the surface, depending on the
sign of $q$. We will return to this point below, in section 4.6.
The formula \annmupfq\ is analogous to the formula for the FZZT-ZZ
annulus in the bosonic string as derived in \KutasovFG. It is
interesting that the annulus \Yrew\ depends on $q$ rather than
$|q|$. This indicates that it involves the exchange of the RR
Ishibashi states in the boundary state \bsfzztmup. (Of course, the
NS Ishibashi states can contribute as well.)

Consider now $\psi_-$. The general target space picture discussed
above implies that $\psi_-$ in the UHP describes a configuration
of a brane with charge $q_b={1\over2}$ and strong-coupling flux
$q_{strong}=q-1$ (see figure 3). Thus the disk amplitude and
annulus between the brane and itself are the same as for $\psi_+$.
Meanwhile, the $q$ dependence in this phase comes from the
interaction of the brane and $|q-1|$ charged ZZ branes located at
$x_{ZZ}(q-1)$. As in our large $x$ expression \sliuhp, the power
of $Y$ is given by $2 q_b q_{strong}$.  It is different from
$\psi_+$ because $q_{strong}$ is different, but it can still be
interpreted as an annulus between an FZZT brane and $|q_{strong}|$
ZZ branes. The prefactor $Z_-(\mu,q)$ leads to exponential
enhancement for positive $q$ and exponential suppression for
negative $q$. This is consistent with the change in the number of
ZZ branes.  It is straightforward to extend this discussion to the
LHP.

It is also instructive to understand in this semiclassical
language what happens to the system as we bring the brane
$B_{-1/2}$ from the LHP to the UHP. Suppose we start with ${\rm
Im}\,\, x$ large and negative, where the system is described by
strong-coupling flux $q$ and a brane with charge $-1/2$.
Increasing $x$ through $x=0$ and out to large positive ${\rm
Im}\,\,x$ corresponds to bringing the brane into the strong
coupling region and back out again. Then we find that the system
is described by strong coupling flux $q-1$ and a brane with charge
$+1/2$. For positive $q$ we see a transition, whereby the brane
{\it picks up a ZZ brane} from the strong coupling region and
carries it back out to weak coupling.  For negative $q$ the brane
{\it leaves behind an anti-ZZ brane} and then returns.

Now let us consider the semiclassical limit for negative $\mu$
(the two cut phase of the 0B theory). Here we find
 \eqn\semlibn{\pmatrix{\psi_+ \cr \psi_-}  \sim {e^{-i{4 \over 3 }
 (x^2 + \mu /2)^{3/2}}\over \sqrt{2}(x^2 + \mu/2)^{1/4}}
 \pmatrix{{(-i\sqrt{2})^{q}( x+\sqrt{x^2
 +\mu/2})^ {{1\over 2} +q }}\left(1 + \CO(\mu^{-3/2})\right)\cr
 { (-\mu)^{-q+{1\over 2}}(-i\sqrt{2})^{q-1} (x+\sqrt{x^2 +\mu/2})
 ^{-{1\over 2}+q}
 }\left(1 + \CO(\mu^{-3/2}) \right)
  }}
The substitution $x= i\sqrt{-{\mu\over2}} \sinh \sigma$ (see
section 3) simplifies this to
 \eqn\semlibns{\pmatrix{\psi_+ \cr \psi_-}  \sim {e^{-{4 \over 3 }
 (-\mu /2)^{3/2}\cosh^3 \sigma}\over\sqrt{2\cosh(\sigma)}}\pmatrix{ (-\mu)^{q/2}
 e^{(q+{1\over2})\sigma} \left(1 + \CO(\mu^{-3/2})\right)\cr
 (-\mu)^{-q/2 }
 e^{(q-{1\over2})\sigma} \left(1 + \CO(\mu^{-3/2})\right)
  }}
This parametrization makes it manifest that these asymptotics are
valid for $x$ in both the UHP and the LHP.

The physical interpretation of these asymptotics is as follows.
The leading effect for both branes arises from a disk amplitude
 \eqn\diskmun{ D(x)=-i{4 \over 3 } (x^2 + \mu /2)^{3/2}= -{4
 \over 3} (-\mu /2)^{3/2}\cosh^3 \sigma }
This value and its derivative $y=\partial_x D(x)= -4 i x \sqrt{x^2
+\mu/2}=-i\mu \sinh(2\sigma)$ are as in section 3 and in
\refs{\KlebanovWG,\SeibergNM}.  The corrections to this result
have not yet been computed using worldsheet methods, but we can
read them off from \semlibns. Here, we will see that the new term
$\pm {1 \over 2} V_R|0\rangle$ in \bsfzztmun\ is important.  Let
us start with $\psi_+$. We interpret the log of the
$q$-independent prefactor in \semlibns\
\eqn\annmunff{
 Z_{annulus}(\sigma,\sigma) = -\log(2\cosh\sigma)+\sigma
 }
as the result of an annulus diagram whose two ends are on the FZZT
brane \bsfzztmun. We can understand better the effect of the new
term $\pm {1\over2}V_R|0\rangle$ by decomposing \annmunff\ into
three terms
 \eqn\annmunffrew{
 Z_{annulus}(\sigma,\sigma)=
 \partial_q D^{naive}(\sigma)-{1\over4}\partial_q^2 F
 +Z_{annulus}^{naive}(\sigma,\sigma)
 }
where the derivatives $\partial_q$ are evaluated at $q=0$.  These
terms have the following interpretations:

\lfm{1.} The first term arises from the disk one point function of
the RR vertex operator with the ``naive" boundary state, by which
we mean \bsfzztmun\ without the extra $\pm {1\over2}V_R|0\rangle$
term:
  \eqn\disks{\partial_q D^{naive}(\sigma) = \langle V_R|\sigma,\eta
  =-1\rangle_{+}^{naive}= \sigma }
This answer can be found by expressing the $\phi$ dependence of
the vertex operator as $\phi e^{Q\phi/2} =\lim_{k\to 0} \partial_k
e^{(k + Q/2) \phi}$, and reading off the coefficient of the RR
Ishibashi state with $iP=-k\sqrt{2}$. (The overall normalization
is fixed by comparing with the matrix model answer \semlibns.) As
a check, note that it is odd under $\sigma\to-\sigma$, as expected
since the R component of \bsfzztmun\ is multiplied by
$\sin(2P\sigma)$.

\lfm{2.} The second term in \annmunffrew\ arises from two
insertions of $V_R$ on the sphere
 \eqn\VRs{-\partial_q^2 F=\langle V_R V_R \rangle = \log(-\mu)}
where we have used the semiclassical expansion \Fleadingcalc.

\lfm{3.} Finally, the third term in \annmunffrew\ arises from the
annulus with two ends on the naive boundary state:
 \eqn\annmunffnaive{Z_{annulus}^{naive}(\sigma,\sigma) =
 -\log(\cosh \sigma) -{1 \over 4} \log(-\mu)
 }
As a check, note that this is invariant under $\sigma \to -\sigma$
as expected from the form of the boundary state. We did not
calculate this using the worldsheet, but simply matched with the
answer \annmunff, assuming the decomposition \annmunffrew. In
particular, the term proportional to $\log(-\mu)$ was designed to
cancel the similar term in \VRs.

\medskip

Armed with these terms, we can now understand the $q$ dependence
of $\psi_+$ in \semlibns. Recall that there are no ZZ branes for
negative $\mu$, and instead $q$ represents the coefficient of the
RR vertex operator in the worldsheet theory. Thus the log of the
$q$-dependent prefactor in \semlibns\ arises from the one-point
function of the RR vertex operator in the boundary state
\bsfzztmun. (Insertions of more vertex operators are higher order
in the semiclassical expansion, and contributions without the
boundary state \bsfzztmun\ are part of the closed string partition
function.) This leads to
 \eqn\annmunfq{
 q\,\partial_q D(\sigma)=q \left( \partial_q D^{naive}(\sigma)-{1\over 2}
 \partial_q^2 F\right) = q\left( \sigma + {1\over2}\log
 (-\mu)\right)
 }
where again $\partial_q$ is evaluated at $q=0$.  The first term
arises from a one point function on a disk with the naive boundary
state and the second from a sphere with two insertions of $V_R$ --
one of them from the background flux and the other from the vertex
operator in the boundary state \bsfzztmun. The fact that
\annmunfq\ agrees with \semlibns\ is a nontrivial check of our
corrected boundary state \bsfzztmun.

Having understood the worldsheet interpretation $\psi_+$, now let
us turn to $\psi_-$. From the form of the boundary state, the
leading order term $D(x)$ and the $q$-independent correction
$Z_{annulus}(\sigma,\sigma)$ in $\psi_-$ should be the same as for
$\psi_+$. But the $q$-dependent correction \annmunfq\ should be
different, because this brane is put in a background flux
$q_{strong}=q-1$. It has the same charge $q_b=+{1\over 2}$, so the
insertion of the RR vertex operator on the disk, which is
proportional to $q_b q_{strong}$, gives $ (q-1)\partial_q
D(\sigma)=(q-1)( \sigma + {1\over 2}\log (-\mu))$. Finally, we
have to account for the crucial factor of $Z_- \approx
(-\mu)^{-q+{1\over 2}}$ which corrects the closed-string free
energy as we change $q $ to $q-1$. Putting all this together, we
find that the worldsheet prediction for the leading-order
correction to $\psi_-$ agrees precisely with the asymptotics of
$\psi_-$ in \semlibns. Of course, this was guaranteed to work,
given the identity \psiidentity. Even so, it is still nice to see
how it all fits together with the worldsheet description.

\subsec{$x\sim 1$, large $|\mu|$  asymptotics}

Another interesting limit is large $|\mu|$ but with $x$ of order
one. In this limit, the theory is still semiclassical, but the
worldsheet is strongly coupled and the branes penetrate deep into
the strong-coupling region. The flux and brane charge are always
fluctuating, so the analytic continuation of $x$ between the UHP
and the LHP must be manifestly smooth. Correspondingly, the
effects of Stokes' phenomenon, as well as the charge conjugation
symmetry \psiC, will be explicit in this limit.

For $\mu\to +\infty$, $x\sim 1$, we find
\eqn\lmxomupqnz{\eqalign{
 \left(\matrix{\psi_+\cr\psi_-}\right)
  &\sim
  e^{-i({4\over3}x^3+\mu
  x)}\left(\matrix{(\epsilon(q)\sqrt{2\mu})^{q}(1+\CO(\mu^{-1/2}))\cr
  (\epsilon(q-1)\sqrt{2\mu})^{q-1}Z_-(\mu,q)(1+\CO(\mu^{-1/2}))}\right)\cr
  &\qquad +
 e^{i({4\over3}x^3+\mu
  x)}\left(\matrix{
     (-\epsilon(q+1)\sqrt{2\mu})^{-q-1}Z_+(\mu,q)
    (1+\CO(\mu^{-1/2}))\cr
   (-\epsilon(q)\sqrt{2\mu})^{-q}(1+\CO(\mu^{-1/2}))}\right)
  }}
As expected, this formula is smooth and is valid for $x$ in both
the UHP and the LHP. Along the real $x$ axis the exact answer has
an anti-Stokes' line, which becomes a branch cut discontinuity in
the classical limit. On the anti-Stokes' line, the two terms in
\lmxomupqnz\ are equally important. Moving $x$ into the UHP (LHP),
the first (second) term dominates as $|x|\to \infty$, as needed
for consistency with the semiclassical limit \sliuhpqu. The
relative coefficient between the two terms is determined by the
charge conjugation symmetry \psiC.

It is interesting to note that a similar phenomenon happens as
$\mu\to +\infty$, due to the factors of $Z_\pm$ in \lmxomupqnz.
(Recall the asymptotic expansions in section 2.2.) That is, each
of the functions $\psi_\pm$ is dominated at large $\mu$ by only
one of the two exponentials in \lmxomupqnz. The other one is
exponentially suppressed in the $\mu \to +\infty$ limit. Which
exponential dominates in $\psi_\pm$ depends on $q$. For $q=0$ we
have $\psi_\pm \sim e^{\mp i({4\over3}x^3+\mu  x)}$. This means
that in this case the fixed charge branes are approximately
``left-moving'' and ``right-moving'' waves. Meanwhile, for $q\ne
0$ we have $\psi_\pm \sim e^{-\epsilon(q) i({4\over3}x^3+\mu x)}$.
Here the fixed charge branes are either both ``left-moving" or
both ``right-moving" waves.

In the limit $\mu\to -\infty$, $x\sim 1$, we find
 \eqn\lmxomun{
 \left(\matrix{\psi_+\cr \psi_-}\right) \sim
 {1\over\sqrt{2}}e^{-{\sqrt{2}\over3}(-\mu)^{3/2}+\sqrt{2}
 \,x^2(-\mu)^{1/2}}\left(\matrix{
 (-\mu)^{q/2}\left(1+\CO(\mu^{-1/2})\right)
 \cr
 (-\mu)^{-q/2}\left(1+\CO(\mu^{-1/2})\right)}\right)
 }
for $x$ in both the UHP and the LHP. One can check that \lmxomun\
agrees exactly with the $x\to 0$ limit of the semiclassical
asymptotics \semlibn-\semlibns. In contrast with the positive
$\mu$ limit \lmxomupqnz, there is only one term (and no
anti-Stokes' line) in the negative $\mu$ limit. This agrees with
the fact that there is no branch cut near $x=0$ in the
semiclassical answer \semlibn.  One can also see that \lmxomun\
satisfies the charge conjugation symmetry. Since \lmxomun\ matches
smoothly onto the semiclassical answer which decays as $x\to \pm
i\infty$, this is evidence that the physical boundary conditions
on $\psi_\pm$ are consistent for every value of $q$.

\subsec{A large $q$ limit}

Finally, let us consider a modified version of the semiclassical
limit in section 4.4, where we not only take $|x|$ and $|\mu|$ to
infinity keeping $\tilde x$ in \txdef\ fixed, but also send
$|q|\to \infty$ keeping
\eqn\qgsfixed{
\tilde q = {q\over \mu^{3/2}} ={\rm fixed}
 }
This limit was studied in \KlebanovWG, where it was shown to
smooth out the Gross-Witten phase transition even in the classical
limit.

Rather than extract the detailed asymptotics of $\psi_\pm$ as in
the preceding subsections, let us just focus on the WKB exponent
for simplicity. This can be efficiently extracted from the
differential equations \flco\ as follows. Take the matrix $A_x$ in
\flco, and drop all derivatives of $r(\mu,q)$. But keep the
derivatives of $\beta(\mu,q)$ -- according to \plII, these are
proportional to $q\sim {1\over g_s}$. The eigenvalues of this
matrix are given by $\pm y(x)$ where
 \eqn\curverew{ y^2=-16\left(x-{i q\over2r^2}\right)^2
 \left(x-{r\over\sqrt{2}}+{i q\over2r^2}\right)
 \left(x+{r\over\sqrt{2}}+{i q\over2r^2}\right) }
(compare with the discussion of the algebraic curve in section 3)
and $r$ satisfies the genus zero version of the string equation
\plIIq, i.e.
 \eqn\streqngz{ - \mu r - r^3 + { q^2 \over r^3}=0 } The WKB
exponents (or at least their $x$-dependent parts) are then
\eqn\WKBexp{ D(x) = \int^x y(x')dx' } where the sign of $y$ is
determined by the boundary conditions on $\psi_\pm$. As a check of
these formulas, note that in the small $\tilde q$ limit, \WKBexp\
reduces to
 \eqn\curvediskzB{
  D(x) = \cases{ -i\left({4\over3}x^3+\mu x\right)+
  q\log\left(x+{i\over2}\epsilon(q)\sqrt{\mu}\right)
  +\CO( \mu^{-3/2}) & $\mu>0$\cr
 -i{4\over3}(x^2+\mu/2)^{3/2}+{ q}\log
\left(x+\sqrt{x^2+\mu/2}\right) +\CO( \mu^{-3/2})& $\mu<0$
 }}
Here we have assumed $x\in{\rm UHP}$, and we have imposed the
proper boundary conditions. \curvediskzB\ agrees with the large
$q$, $x$ dependent part of the semiclassical expansions \sliuhpqu\
and \semlibn, showing that the various limits are consistent with
one another.

The algebraic curve $y(x)$ in \curverew\ defines a Riemann surface
that is a double cover of the complex $x$ plane. It generalizes
the $q=0$ surface of the $\eta=-1$ FZZT brane reviewed in section
3. In \SeibergNM, it was shown how this surface unifies
geometrically many features of minimal string theory. So let us
examine its $q$-deformed generalization in more detail. According
to \streqngz\ (see also \solqab\ and \solqa), $r$ is nonzero when
$q$ is nonzero. Then the curve \curverew\ always has exactly one
singularity at \eqn\singcurveq{ x={i q\over2r^2} } for either sign
of $\mu$. For $\mu>0$, the $q=0$ curve \diskymup\ has two
singularities at $x=\pm {i\sqrt{\mu}\over2}$, so nonzero $q$ has
the effect of opening up one of these singularities while shifting
continuously the other. For $\mu<0$, the $q=0$ curve \diskymun\
has only one singularity at $x=0$, and one can check using
\solqab\ that nonzero $q$ continuously shifts the location of this
singularity without opening it up.

Physically, changing $q$ corresponds to adding charged ZZ branes
to the system for $\mu>0$, and it corresponds to insertions of the
closed-string RR vertex operator for $\mu<0$. Therefore, our
analysis of the $q$-deformed curve is consistent with the general
idea in \KutasovFG\ that adding background ZZ branes opens up
singularities, while deforming by closed string operators
preserves the singularities.

Let us take a closer look at the $q$ deformation for $\mu>0$. When
$q$ is zero, the singularities are in the upper and lower half
plane, symmetric under $x\to -x$. This symmetry is broken by
nonzero $q$. Turning on $q$ splits one of the singularities
depending on the sign of $q$. According to \singcurveq, the split
singularity is in the LHP (UHP) for $q>0$ ($q<0$). We identify the
sign of $q$ (and which singularity is split) with the charge of
the $(1,1)$ ZZ brane that exists in the phase.  Note that we
already saw a hint of this symmetry breaking in our study of the
annulus amplitude \annmupfq.

Another interesting feature of the $q$ deformed curve can be seen
by expanding $y(x)$ around $x=\infty$. From \curverew, we find
(for both signs of $\mu$)
\eqn\curverewexp{
y = -i(4x^2+\mu)+{q\over x}+\dots
 }
Therefore, we can characterize the effect of $q$ as a deformation
of the curve which introduces a pole at infinity. Equivalently, we
can say that the Riemann surface of the charged brane has a
puncture at $x=\infty$. Later, when we study the surface of the
neutral brane, we will see that this feature at infinity has an
important role to play.

\newsec{The neutral branes}

\subsec{The differential equations}

In addition to charged branes, our system also has a stable
neutral brane $B_0$ in its spectrum, corresponding to the
$\eta=+1$ FZZT brane in the worldsheet description. This brane is
most natural in the 0A language. Using the dual matrix model, one
finds that the partition function
\eqn\neutralpartfn{
\psi_0=\psi_0(x,\mu,q)={\langle
B_0(x)\rangle_{\mu,q}\over\CZ(\mu,q)}
}
satisfies the following differential equations:
\eqn\bakercl{\eqalign{
 &\partial_\mu^2\psi_0 = (r^2+\mu-2x^2)\psi_0\cr
 &x\partial_x\psi_0 =
 (rr'-|q|)\psi_0-(r^2+4x^2)\partial_\mu\psi_0
 }}
Again, the derivation of \bakercl\ is left to appendix A.2.  The
structure of these equations is similar to the corresponding
equations in the bosonic string \refs{\MooreMG,\MooreCN}. As for
the charged branes, we can list several consistency checks. First,
the equations are clearly invariant under the shift symmetry of
$\beta$ (recall the discussion below \qmqsym), so $\psi_0$ indeed
has zero charge. Second, one can check that the compatibility of
the two equations in \bakercl\ is equivalent to the string
equation \plIIq. Finally, one can show using the results of
\FlaschkaWJ\ that there exists a solution to \bakercl\ that is
smooth in the entire complex $x$ plane.

Given that $\psi_0$ is the double-scaling limit of the orthonormal
wavefunction $\psi_N = e^{-V(x^2)/2}P_N(x^2)$ of the complex
matrix model (see appendix A.3), it must have the following
symmetries:
\eqn\bakersymm{
 \psi_0(x,\mu,q) = \psi_0(-x,\mu,q),\qquad \psi_0(x,\mu,q)=\psi_0(x,\mu,-q)
 }
In addition, we expect that $\psi_0 \to 0$ exponentially as $ix\to
\pm \infty$, since the matrix model potential goes to infinity
there.

As for the charged brane, there exists an identity which relates
branes of different $q$ (see e.g.\ \ItsKapaev):
\eqn\neutralbraneid{
 \psi_0(x,\mu,q+1) = {1\over x^2}\big(\partial_\mu\psi_0(x,\mu,q)-
 \psi_0(x,\mu,q)\partial_\mu \log Z_+(\mu,q)\big)
}
Here we have assumed $q\ge 0$ for simplicity; the result for $q<0$
can be obtained using the charge conjugation symmetry \bakersymm.
It is straightforward to check explicitly that \neutralbraneid\
satisfies the differential equations \bakercl\ with $q\to q+1$. In
appendix A.3, we also offer a microscopic derivation starting from
the complex matrix model. Using the definition \neutralpartfn\ of
$\psi_0$, we can also write \neutralbraneid\ as
\eqn\neutralbraneidii{
 \langle B_0(x)\rangle_{\mu,q+1} = {1\over
 x^2 \CZ(\mu,q)}\Big(\langle B_0(0)\rangle_{\mu,q}\,\partial_\mu \langle B_0(x)\rangle_{\mu,q} -
 \partial_\mu\langle B_0(0)\rangle_{\mu,q}\,\langle
 B_0(x)\rangle_{\mu,q}\Big)
}
Here we have also used $\psi_0(x=0,\mu,q) = Z_+(\mu,q)$ for $q>0$,
a fact that we will discuss further in the next subsection and
derive in appendix A.3. There, we will also interpret
\neutralbraneidii\ as the statement that for the neutral brane,
changing $q\to q+1$ is equivalent to inserting $B_0$ at $x=0$.

\subsec{Asymptotics}

The various asymptotic limits of the neutral brane are much
simpler than than those of the charged branes.

\lfm{1.} {\it Large $x$, everything else held fixed.} Then
\eqn\bakerxl{
\psi_0 \sim {1\over
2\sqrt{\pi}}\left(-{ix\over\sqrt{2}}\right)^{-|q|-1/2}e^{-i\sqrt{2}({4\over3}x^3-\mu
x)}\left(1+\CO(1/x)\right)
}
in the UHP, and similarly in the LHP using \bakersymm. Note that
these simple asymptotics are valid for any sign of $\mu$, and for
both $q$ zero and nonzero.

\lfm{2.} {\it Semiclassical limit, $\mu<0$}. The string equation
is solved by $r=r(\mu,q)$ given in \solqab, and we have
 \eqn\wkbpsifin{ \psi_0 \sim
{1\over
2\sqrt{\pi}}\left(-{ix\over\sqrt{2}}\right)^{-|q|-1/2}e^{-i\sqrt{2}({4\over3}x^{3}-\mu
 x)}(1+\CO(\mu^{-3/2})) }
in the UHP. This answer obviously agrees with \bakerxl\ at $|x|\to
\infty$. The overall factor of $x^{-|q|}$ can be removed by
redefining $\psi_0$.  If we do that, it will change the term $|q|$
in \bakercl.  Our choice of \bakercl\ and $\psi_0$ guarantees that
$\psi_0$ is entire in $x$ and for all (not necessarily integer)
$q$. Apart from this overall factor, we see from \wkbpsifin\ that
$q$ does not contribute to either the disk or the annulus
amplitude in this phase. We can understand this as follows. In
this phase, we have flux but no charged ZZ branes. Since the
neutral brane has no RR Ishibashi component in its boundary state,
the disk with a single insertion of the RR vertex operator must be
zero. Thus the leading order semiclassical limit is necessarily
$q$-independent. The first contribution will come at $\CO(q^2)$,
from the two-point function of the RR vertex operator.

\lfm{3.} {\it Semiclassical limit, $\mu>0$}. Here $r(\mu,q)$ is
given by \solqa\ for $q\ne 0$ and by \claplII\ for $q=0$. However,
we find a unified formula for the asymptotics
\eqn\wkbpsifiniii{
\psi_0 \sim
 {1\over
\sqrt{2\pi}}\left({\mu}-2x^2\right)^{-1/4}\left({\sqrt{\mu}+\sqrt{{\mu}-2x^2}\over
2}\right)^{-|q|}
 e^{-{4\over3}\sqrt{2}({\mu\over2}-x^2)^{3/2}}(1+\CO(\mu^{-3/2}))
}
in the UHP. The answer in the LHP is similar, using \bakersymm.
Again, one can check that this correctly reduces to \bakerxl\ as
$x\to +i\infty$. The fact that $\psi_0$ depends only on $|q|$
implies that its interaction with the charged ZZ branes that exist
in this phase proceeds solely through the exchange of NSNS states.
This is consistent with the boundary state for the neutral brane,
which does not have a RR Ishibashi component.

\lfm{4.} {\it Small $x$, everything else held fixed.} Then we find
 \eqn\bakerxz{ \psi_0\sim Z_+(\mu,|q|)(1+\CO(x^2)) }
where $Z_\pm=re^{\mp\beta}$ as above. The answer clearly respects
the symmetries \bakersymm. It agrees with the small $x$ limit of
the semiclassical limit for $\mu>0$, but not for $\mu<0$. Of
course, the differential equations \bakercl\ only fix $\psi_0$ up
to an overall $q$-dependent factor. More work is required to show
that $\psi_0$ reduces to precisely $Z_+(\mu,|q|)$ at $x=0$ with no
additional $q$-dependent normalization factor. We show this in
appendix A.3.

\subsec{A large $q$ limit}

Just as for the charged branes, it is interesting to consider a
modified semiclassical limit, where we take $q$ to infinity
holding fixed $\tilde q = q/\mu^{3/2}$. To extract the WKB
exponent from the differential equations \bakercl, we define the
column vector $\Psi_0=\left(\matrix{\psi_0\cr
\partial_\mu\psi_0}\right)$. The action of $\partial_x$ on
$\Psi_0$ is a $2\times 2$ matrix
\eqn\QPonPsi{
 \partial_x\Psi_0 = {1\over x}\left(\matrix{ r r'-|q| &
  -4x^2-r^2 \cr
  8x^4-2x^2(2{\mu}+r^2)+(r'^2-{q^2\over r^2})
   & -rr'-|q|}\right)\Psi
}
After dropping the $\mu$ derivatives, the eigenvalues of this
matrix become $-{|q|\over x}\pm y(x)$ with $y(x)$ satisfying
\eqn\curvefactorzA{
y^2 = -{(2x^2-r^2-{\mu})(4x^2+r^2)^2\over x^2}
}
Here $r$ again satisfies the genus zero string equation \streqngz.
The WKB exponent of $\psi_0$ is then
\eqn\wkbneut{
D(x) = -|q|\log x + \int^x y(x')dx'
}
As a consistency check, we can compare the small $q$ limit of
\wkbneut\ with the large $q$ limit of the semiclassical limit
above. From \wkbneut, we find
 \eqn\curvedisk{
D(x) = \cases{  -{4\over3}\sqrt{2}({\mu\over2}-x^2)^{3/2}
-|q|\log\left(\sqrt{\mu\over2}+\sqrt{{\mu\over2}-x^2}\right)
+\CO(q^2) & $\mu > 0$ \cr
 -i\sqrt{2}({4\over3}x^3-\mu x) -|q|\log x+
\CO(q^2) & $\mu<0$\cr}
 }
which agrees precisely with the asymptotics
\wkbpsifin--\wkbpsifiniii.

The algebraic curve \curvefactorzA\ generalizes to nonzero $q$ the
curve of the $\eta=+1$ FZZT brane reviewed in section 3 and
studied in detail in \SeibergNM. For nonzero $q$, the curve
\curvefactorzA\ always has two singularities at $x=\pm i r/2$ and
a pole at $x=0$ with residue $\pm q$.  For $\mu<0$, the two
singularities at $x=\pm i r/2$ are continuously connected to the
two singularities at $x=\pm i \sqrt{-\mu}/2$ of the $q=0$ surface
\diskymun.  So the main effect of turning on $q$ is to create the
pole at $x=0$. For $\mu>0$, the singularity at $x=0$ of the $q=0$
surface \diskymup\ is split to two singularities at $x=\pm i r/2$,
and a pole at $x=0$ is created. The $q$ deformation in this phase
comes from charged ZZ branes; specifically, we expect it to be the
result of the annulus diagram between the $\eta=+1$ FZZT and the
$\eta=-1$ ZZ brane.

As we noted in section 3, the $q=0$ curves for the two signs of
$\eta$ are essentially the same, related by $\mu\to -\mu$.
However, comparing \curverew\ and \curvefactorzA\ shows that they
are quite different for nonzero $q$. There are two important
differences between them.  First, for the neutral brane $y(x)$
\curvefactorzA\ has a pole with residue $\pm q$ at $x=0$, while
for the charged brane \curverew\ $y(x)$ is smooth at every finite
$x$. Second, the charged brane curve has such a pole at infinity
\curverewexp, while the curve of the neutral brane does not have a
pole there:
 \eqn\curvezalgx{ y =
 -i\sqrt{2}(4x^2-\mu)+\CO\left({1\over x^2}\right) }
Equivalently, we can say that the surface for the charged brane
has a puncture at $x=\infty$, while the surface for the neutral
brane has a puncture at $x=0$. These differences serve to
distinguish between the surfaces of the charged and neutral
branes, which would otherwise be related by $\mu\to -\mu$.

\vskip 0.8cm

\noindent {\bf Acknowledgments:}

We would like to thank D.~Gaiotto, S.~Hellerman, A.~Its,
V.~Kazakov, I.~Klebanov, J.~Maldacena, G.~Moore and K.~Okuyama for
useful discussions. NS thanks his friends and colleagues at the
Racah Institute of Physics for hospitality during the completion
of this work.  The research of NS is supported in part by DOE
grant DE-FG02-90ER40542. The research of DS is supported in part
by an NSF Graduate Research Fellowship and by NSF grant
PHY-0243680. Any opinions, findings, and conclusions or
recommendations expressed in this material are those of the
author(s) and do not necessarily reflect the views of the National
Science Foundation.

\vfill

\appendix{A}{The integrable hierarchies of the minimal superstring}

\subsec{Type 0B}

The matrix model dual to 0B minimal string theory is \KlebanovWG:
\eqn\mmdefzb{
\CZ_N = \int dM e^{-\Tr V(M)}
}
where $M$ is an $N\times N$ Hermitian matrix, and $V(M)$ is
polynomial in $M$. The model describing our system has $V(M)=-
M^2+g M^4$; the continuum ``double-scaling" limit consists of
taking $N\to \infty$ while simultaneously tuning $g$ to some
critical value. In this limit, $\CZ_N$ becomes the bulk partition
function $\CZ(\mu,q)$, and the eigenvalues of $M$ become
distributed along finite ``cuts" on the real axis.

Recall that in the bosonic string, the FZZT partition function
$\psi_{bos}(x)$ is obtained from double-scaling the orthonormal
wavefunction of the matrix model,
\eqn\bosfzztpart{
 \psi_{bos}(x) = \lim_{N\to \infty} \psi_N(x)
 }
which at finite $N$ is the expectation value of the determinant
operator $\det(x-M)$,
\eqn\bosfzztpartN{
 \psi_N(x)=  {1\over\sqrt{h_N}}e^{-V(x)/2}\langle
 \det(x-M)\rangle
 }
Here the story is similar, except that the even and odd
orthonormal wavefunctions have different scaling limits. That is,
if we take $N$ to be even, then we study the two functions
$\psi_{even}(x)$ and $\psi_{odd}(x)$ via
\eqn\evenoddfzztpart{\eqalign{
 &\psi_{even}(x) = \lim_{N\to \infty}\psi_N(x)\cr
 &\psi_{odd}(x) = \lim_{N\to \infty}\psi_{N-1}(x)
 }}
$\psi_N(x)$ is still related to the determinant operator through
\bosfzztpartN, while $\psi_{N-1}(x)$ is related to the {\it
inverse determinant} through
\eqn\oddfzztpartN{
 {1\over \sqrt{h_{N-1}}}e^{V(x)/2}\left\langle {1\over
 \det(x-M)}\right\rangle =
  \int_{-\infty}^{\infty} d\lambda\,e^{(V(x)-V(\lambda))/2}
  { \psi_{N-1}(\lambda)\over
  x-\lambda}
 }
$\psi_{N-1}$ can be extracted from this expectation value by
subtracting the value of this function above and below the real
axis.  This makes it clear that both $\psi_{even}$ and
$\psi_{odd}$ are physical observables of the matrix model.

Now, we claim that the partition functions of the charged FZZT
branes correspond to the linear combinations \eqn\chargedlincombo{
\psi_\pm(x) = \psi_{even} (x)\pm i \psi_{odd}(x) } To see this, we
need to quickly review the Zakharov-Shabat integrable hierarchy
which underlies 0B minimal superstring theory
\refs{\PeriwalGF\NappiBI\CrnkovicMS\CrnkovicWD\HollowoodXQ-\BrowerMN}.
It can be formulated in terms of $2\times 2$ matrix differential
operators\foot{We thank D.\ Gaiotto for help with the operator
formalisms described in this subsection and the next.}
 \eqn\Qdef{
 Q = \left(\matrix{ i \partial_\tau & -{iZ_+\over\sqrt{2}} \cr {i
 Z_-\over\sqrt{2}} & -i\partial_\tau}\right) } and \eqn\Pdef{ P =
 \sum_{j\ge 0}\left( t_{2j+1}(-Q^2)^{j+1/2}_+ +
 t_{2j}\left(Q(-Q^2)^{j-1/2}\right)_+\right) }
which must satisfy the string equation \eqn\ZSstreqn{ [P,Q]=1 }
Here $\tau=t_0$ is the coupling to the lowest dimension operator.

The defining property of $\psi_\pm$ is that they are the
Baker-Akhiezer functions of the ZS integrable hierarchy. That is,
they satisfy the differential equations
\eqn\psipmdefzb{
Q\left(\matrix{\psi_+\cr \psi_-}\right) = x\left(\matrix{\psi_+\cr
\psi_-}\right),\qquad P\left(\matrix{\psi_+\cr \psi_-}\right) =
-\partial_x\left(\matrix{\psi_+\cr \psi_-}\right)
}
Comparing with \Qdef, we see that if $Z_\pm$ have charge $\pm 1$
under the RR shift symmetry $\beta\to \beta+const$ (recall the
discussion under \qmqsym), then $\psi_\pm$ have definite charge
$\pm 1/2$. This shows that these are indeed the partition
functions of the charged FZZT branes.\foot{When checking
\psipmdefzb\ against \ZSstreqn, one should keep in mind that since
$P$ and $Q$ are differential operators in $\mu$, they commute with
with $x$ and $\partial_x$. Thus $PQ\Psi = P(x\Psi) = x P \Psi =
-x\partial_x\Psi$, and similarly for $QP\Psi$.}

For the special case of $(p,q)=(2,4)$ relevant to the paper, we
set $t_2=-4$, $\tau=t_0=\mu$, and all other $t_j=0$. Then
\eqn\Qzbagain{
Q=\left(\matrix{i\partial_\mu & -{i Z_+\over\sqrt{2}}\cr {i
Z_-\over\sqrt{2}} & -i\partial_\mu}\right)
}
and the general formula for $P$ yields
\eqn\Pfind{
 P = -4\left(Q (-Q^2)^{1/2}\right)_+ + \mu \left(Q (-Q^2)^{-1/2}\right)_+=
 -i\pmatrix{4\partial_\mu^2-r^2 -\mu& -4{Z_+ \over\sqrt{2}}\partial_\mu - \sqrt{2}Z_+' \cr
 4{Z_-\over\sqrt{2}}\partial_\mu +{\sqrt{2}Z_-'}  & -4\partial_\mu^2+r^2+\mu }
 }
One can check that $[P,Q]=1$ is equivalent to the string equation
\plII. It is also straightforward to check that the defining
equations \psipmdefzb\ are the same as the differential equations
\flco\ studied in the paper.

\subsec{Type 0A}

The matrix model dual to 0A minimal string theory is \KlebanovWG
 \eqn\mmdefza{ \tilde\CZ_N(q) = \int dMdM^\dagger\, e^{-\Tr V(M
 M^\dagger)} }
where $M$ is a $(N+|q|)\times N$ complex matrix, and $V( M
M^\dagger)$ is polynomial in $M M^\dagger$. By bringing $M$ to the
form $M_{ij}=\sqrt{\lambda_i} \delta_{ij}$, we can reduce
\mmdefzb\ down to an integration over eigenvalues: \eqn\mmzaeigen{
\tilde\CZ_N(q) = \int_{0}^{\infty}d\lambda_i\, \lambda_i^{|q|}
e^{-V(\lambda_i)} \Delta(\lambda)^2 } The model describing our
system has $V(\lambda)= -\lambda + \tilde g \lambda^2$; the
continuum limit consists of taking $N\to \infty$ while
simultaneously tuning $\tilde g$ to its critical value. In this
limit, $\tilde\CZ_N$ becomes the bulk partition function
$\CZ(\mu,q)$ (the same as the 0B partition function, in this
special case), and the eigenvalues $\lambda_i$ become distributed
on a single cut ending at the origin.

The partition function $\psi_0$ of the neutral FZZT brane is
obtained from the double-scaling limit of the nearly orthonormal
wavefunction
\eqn\detevza{
\psi_N(\lambda,q) ={1\over\sqrt{h_N}}e^{-V(\lambda)/2}
\langle\det(\lambda-MM^\dagger) \rangle
}
By nearly orthonormal, we mean that the wavefunctions satisfy
\eqn\nearlyorth{
\int_{0}^{\infty} d\lambda\,
\lambda^{|q|}\psi_m(\lambda,q)\psi_n(\lambda,q) = \delta_{m,n}
}
In this limit
\eqn\orthwvfnlim{
\psi_0(x,q)=\lim_{N\to \infty}\psi_N(x^2,q)
}
is the Baker-Akhiezer function of the integrable hierarchy
underlying the 0A minimal string. This integrable hierarchy
belongs to the usual KP hierarchy of the bosonic string, but with
a few important differences
\refs{\MorrisBW\DalleyQG\DalleyBR\LafranceWY-\DiFrancescoRU,\JohnsonHY}.
Instead of Lax operators $P$ and $Q$, the operators here are
 \eqn\Qdefza{ Q = \partial_\tau^2-u
 }
($\tau=t_0$ is again the coupling to the lowest-dimension
operator, and $u=\partial_\tau^2\log \tilde\CZ$ is the specific
heat) and
 \eqn\Ddefza{ D = -\sum_{j\ge
 0}(j+{1\over 2})\,t_{2j}\, Q^{j+1/2}_+
 }
with $Q$ and $D$ satisfying the string equation
 \eqn\zeroAstreqn{
 [D,Q]=Q
 }
The defining equations for $\psi_0$ are then
 \eqn\psizdefza{
 Q\psi_0 = -2x^2\psi_0,\qquad D\psi_0 =
 \left(-{1\over2}x\partial_x-{1\over2}|q|-{1\over4}\right)\psi_0
 }
Since $Q$ and $D$ are invariant under $\beta\to \beta+const$,
$\psi_0$ is indeed neutral under the RR symmetry.

For $(p,q)=(2,4)$, we set $\tau=t_0=\mu$ and $t_2=2/3$. Then the
Lax operators are
\eqn\DQtwofour{
Q = \partial_\mu^2-u,\qquad D = -{1\over2}\mu
\partial_\mu-(\partial_\mu^3-{3\over4}\{u,\partial_\mu\})
 }
One can check that with $u=r^2+\mu$, $[D,Q]=Q$ becomes equivalent
to the string equation \plIIq\ after multiplying the equation by
$r^2$ and integrating once with respect to $\mu$.  By taking
appropriate linear combinations of the two equations in
\psizdefza, we can reduce them down to the form \bakercl.

\subsec{More on the neutral brane in the complex matrix model}

Finally, let us analyze in more detail the matrix model
description of the neutral brane, starting from \detevza. This is
useful for a variety of reasons. First, we will see how the
identity \neutralbraneid\ can be easily derived using the matrix
model. Second, we will confirm the symmetries and boundary
conditions imposed at the end of section 5.1, as well as the
asymptotics derived in section 5.2. For simplicity, we will assume
that $q\in \Bbb Z$.

To begin, let us note that $|q|$ can be increased to $|q|+1$ in
the complex matrix model by inserting a determinant operator
$\det(\lambda-M M^\dagger)$ at $\lambda=0$, since this has the
effect of changing the measure from $\lambda^{|q|}$ to
$\lambda^{|q|+1}$. Therefore, starting from the finite $N$ formula
\detevza\ in background $q$, we find that
\eqn\detevzaqp{
 \psi_N(\lambda,|q|+1) = {1\over\sqrt{h_N}} e^{-V(\lambda)/2}
\big\langle \det(-MM^\dagger)\big\rangle_{|q|}^{-1}
\big\langle\det(\lambda-MM^\dagger)\,
\det(-MM^\dagger)\big\rangle_{|q|}
}
Here the expectation value is evaluated in background $|q|$. Using
the determinant formulas in \MorozovHH, this becomes
\eqn\detevzaqpii{
 \psi_N(\lambda,|q|+1) \sim
 {1\over\lambda}\psi_N(0,|q|)^{-1}\big(\psi_{N+1}(\lambda,|q|)\psi_N(0,|q|)
 -\psi_N(\lambda,|q|)\psi_{N+1}(0,|q|)\big)
}
The reason we can apply the determinant formulas of \MorozovHH\ is
because they depend only on the orthogonality of the polynomials
and not on the details of the measure or limits of integration.

In the double scaling limit, the shift in the index of the
orthonormal wavefunction becomes a derivative with respect to
$\mu$. Therefore, \detevzaqpii\ becomes
\eqn\detevzaqpiii{
 \psi_N(x^2,|q|+1) \to \psi_0(x,\mu,|q|+1) =
 {1\over x^2}\Big(\partial_\mu\psi_0(x,\mu,|q|)
 -\psi_0(x,\mu,|q|)\partial_\mu\log \psi_{0}(0,\mu,|q|)\Big)
}
Finally, using the fact that $\psi_0(0,\mu,|q|)=Z_+(\mu,|q|)$ (see
\bakerxz, and also the calculations below), we arrive at precisely
the identity \neutralbraneid\ in the text.

Although \neutralbraneid\ is useful for relating $|q|$ to $|q|+1$,
a more straightforward way to understand the $q$ dependence of the
neutral brane is to relate everything directly to $q=0$. To do
this, note that another way of writing \detevza\ is:
\eqn\psiqza{
\psi_N(\lambda,q)= {1\over\sqrt{h_N}}e^{-V(\lambda)/2}\left\langle
\big(\det(-M M^\dagger)\big)^{|q|}\right\rangle^{-1}{\left\langle
\det(\lambda-MM^\dagger )\big(\det(-M
M^\dagger)\big)^{|q|}\right\rangle}
}
where now $\langle \,\, \rangle$ denotes the expectation value
taken in a background of $q=0$. Using the determinant formulas in
\MorozovHH, this becomes
\eqn\psiqzabec{
\psi_N(\lambda,q)= \left({\det_{k,l}\,
\psi_{N+k-1}(\lambda_l,0)\over \det_{k,l}\,\lambda_l^{k-1}
}\right)^{-1}\left({\det_{i,j}\, \psi_{N+i-1}(\lambda_j,0)\over
\det_{i,j}\,\lambda_j^{i-1}}\right)
}
where $1\le i,\,j\le |q|+1$; $1\le k,\,l\le |q|$;
$\lambda_1=\dots=\lambda_{|q|}=0$; and $\lambda_{|q|+1}=\lambda$.
Therefore, in the double-scaling limit \psiqzabec\ becomes
\eqn\psiqzaii{
\psi_N(x^2,q)\to \psi_0(x,\mu,q)=
  \left({ \det_{k,l}\,\partial_\mu^{l-1}\psi_0(\sqrt{\lambda_k},\mu,q=0)\over
   \det_{k,l}\,\lambda_k^{l-1} }\right)^{-1}\left( { \det_{i,j}\,\partial_\mu^{j-1}\psi_0(\sqrt{\lambda_i},\mu,q=0)\over
   \det_{i,j}\,\lambda_i^{j-1} }\right)
}
Here $i$, $j$, $k$, $l$ are as in \psiqzabec;
$\lambda_1=\dots=\lambda_{|q|}=0$; and $\lambda_{|q|+1}=x^2$. The
limit to $\lambda_1=\dots=\lambda_{|q|}=0$ is slightly subtle --
it involves the fact (see \bakerxz) that $\psi_0$ depends only on
$x^2$ around $x=0$.

Eq.\ \psiqzaii\ is a completely general formula for the partition
function of the neutral brane, valid for all $(2,2k)$ 0A
multicritical points. Since the only input is the neutral brane
partition function at $q=0$, we can use \psiqzaii\ to learn about
the system at nonzero $q$, at least in principle. This can be
difficult in practice, however, since the formula in its full
generality is rather complicated. Fortunately, it simplifies and
becomes more useful in the various limits studied in section 5.1.

For instance, a straightforward calculation shows that in the
semiclassical limit, \psiqzaii\ reduces to
 \eqn\psiqzasemiconclude{
 \psi_0(x,\mu,q) \approx
 \big(z(x,\mu)-z(0,\mu)\big)^{|q|}x^{-2|q|}\psi_0(x,\mu,q=0)
 }
where
\eqn\zdefapp{
z(x,\mu)\equiv \partial_\mu D(x,\mu)
}
and $D(x,\mu)$ is the WKB exponent of $\psi_0$ at $q=0$. To derive
\psiqzasemiconclude, use the fact that $\psi_0 \approx
e^{D(x,\mu)+\dots}$ in the semiclassical limit, implying that
\eqn\dpsizdmuapprox{
\partial_\mu^{m}\psi_0 \approx z^{m}\psi_0
}
to leading order. Again, let us emphasize that
\psiqzasemiconclude\ is completely general, valid for all $(2,2k)$
0A models. The only input is $D(x,\mu)$, which can be derived from
the boundary state of the $q=0$, $\eta=+1$ FZZT brane. For the
case of $(2,4)$ considered in this paper, our results in section 3
imply
 \eqn\zofxza{
 z(x,\mu) = \cases{ \sqrt{2}ix & $\mu<0$\cr
               -\sqrt{2}\sqrt{{\mu\over2}-x^2} & $\mu>0$\cr}
 }
Substituting this into \psiqzasemiconclude, we reproduce precisely
the $q$-dependence of the asymptotics \wkbpsifin\ and
\wkbpsifiniii.

It is also just as straightforward to compare the general matrix
model formula \psiqzaii\ with the small $x$ limit \bakerxz. In
this limit, we find the simple result from the matrix model:
\eqn\smallxmat{
\psi_0(x,\mu,q) \to {\tilde\CZ(\mu,|q|+1)\over \tilde\CZ(\mu,|q|)}
}
This formula is general, valid for all $(2,2k)$. For the special
case of $(2,4)$, we have $\tilde\CZ(\mu,q)=\CZ(\mu,q)$. And for 0B
we have the identity \ZFidentity, $\CZ(\mu,q+1) =
\CZ(\mu,q)Z_+(\mu,q)$. When substituted into \smallxmat, this
gives precisely the small $x$ asymptotics \bakerxz\ which are
smooth at $x=0$.

\appendix{B}{Proving some useful identities}

\subsec{Closed-string sector}

In this appendix, we will analyze more carefully various
identities used in the text. Let us start with the closed-string
sector and explore in more detail the properties of the string
equation \plII. In particular, we wish to justify the treatment of
the integration constants alluded to in footnote 2. The issue is
that since the string equations \plII\ depend only on $\beta'$,
the function $\beta$ is determined up to an additive $q$ dependent
integration constant. This ambiguity amounts to freedom to rescale
$Z_\pm(\mu,q)$ by $B(q)^{\pm 1}$ with $B$ an (almost) arbitrary
function of $q$. $B(q)$ is obviously constrained by the charge
conjugation symmetry $Z_+(\mu,q)=Z_-(\mu,-q)$, which implies that
$B(q)=B(-q)^{-1}$, and $B(0)=1$. The question is what other
constraints we must impose on $B(q)$ in order to determine it
completely.

The solutions of the equations for different values of $q$ are
related by\foot{Note that these identities might not be consistent
with the physically acceptable boundary conditions on $r(\mu,q)$
at $\mu\to \pm \infty$. This is discussed more in the text.}
 \eqn\clstridentityapp{
 r(\mu,q\pm 1)^2 =r(\mu,q)^2-
 2\partial_\mu^2\log Z_\pm(\mu,q)
 }
or equivalently by
\eqn\clstridentityappii{
 r(\mu,q\pm 1)^2=
 -r(\mu,q)^2-2\mu+ 2(\partial_\mu\log Z_\pm(\mu,q))^2
 }
The proof of the first identity was sketched in the text, and the
second identity is trivially related to the first through the
definition of $Z_\pm$ \Zpmd\ and the string equation \plIIq. Using
the symmetry $r(\mu,q)=r(\mu,-q)$ in \clstridentityapp, we find
 \eqn\clstridentityappiii{
 \partial_\mu^2
 \log \left(Z_+(\mu,q-1) Z_-(\mu,q)\right) = 0 }
Integrating twice, we find \eqn\ZpZmprodAB{ Z_+(\mu,q-1)Z_-(\mu,q)
= a(q) e^{b(q) \mu} } where $a(q)$ and $b(q)$ are integration
constants. It is evident from the asymptotic expansions in section
2.2 that $b(q)=0$. Then, using the freedom to rescale $Z_{\pm}$ by
$B(q)^{\pm 1}$ we can set $a(q)=1$.  We conclude that
 \eqn\clstridentityappiv{ Z_+(\mu,q-1)Z_-(\mu,q) = 1 }
as claimed in the text. Using \partf\ and the same integration
constants as in \clstridentityappiv, we can also integrate
\clstridentityapp\ twice to conclude
 \eqn\ZFidentityapp{Z_\pm(\mu,q)=e^{F(\mu,q)-F(\mu,q\pm1)}=
 {\CZ(\mu,q\pm 1) \over \CZ(\mu,q)}}
which reproduces \ZFidentity\ in the text.

The identities presented here can be written a number of different
ways, some of which may be more familiar to others. For instance,
if we define
 \eqn\fdef{f_\pm \equiv \partial_\mu\log Z_\pm = {r'\over r} \mp \beta' =
 {r'\over r} \mp {q\over r^2}
 }
then, using \plII\ or \plIIq\ we readily find that they satisfy
another version of the Painleve II equation
 \eqn\feq{f_\pm'' + 2 \mu f_\pm - 2 f_\pm^3  =\pm 2 q+1}
These solutions are related by
 \eqn\fpmid{f_+^2 + f_+'=f_-^2 + f_-'}
Another way of writing \plIIq\ is in terms of the specific heat
$v=F''={1\over 2
} r^2$, which satisfies
 \eqn\ptf{v'' -{(v')^2\over 2 v}- 4 v^2- 2 \mu v + {q^2 \over
 2v}=0}
This is known as the Painlev\'e 34 equation, and $f_\pm = {v'\mp q
\over 2v}$. Finally, in terms of $f_\pm$ the relation
\clstridentityapp\ between the values with different $q$ is known
as the Miura transformation,
\eqn\miurat{
f_\pm(\mu,q)^2 - \partial_\mu f_\pm(\mu,q) = f_\pm(\mu,q\pm 1)^2 +
\partial_\mu f_\pm(\mu,q \pm 1)
}
which is a consequence of \feq.

\subsec{Open-string sector}

Next let us turn to the open-string sector. We wish to sketch a
proof of \psiidentity\ in the text:
 \eqn\psiidentityagain{ \psi_-(x,\mu,q) = Z_-(\mu,q)\psi_+(x,\mu,q-1)
 }
Our strategy for the proof will be to explicitly construct a
solution of the differential equations \flco\ that obeys
\psiidentityagain.

The first step of the proof is define a function $S=S(x,\mu,q)$
via the differential equations
\eqn\diffeqS{\eqalign{
 &0 =\partial_\mu \log S -\Big(2ix-\partial_\mu\log
  Z_-\Big)-{1\over\sqrt{2}}S^{-1}+{1\over\sqrt{2}} r^2
  S\cr
 & 0 = \partial_x\log
 S-2i(4x^2+r^2+\mu)+i\big(2\sqrt{2}ix
 +\sqrt{2}\partial_\mu \log Z_-\big)
 S^{-1}\cr
 &\qquad\qquad\qquad -i\big(2\sqrt{2}ix
 -\sqrt{2}\partial_\mu\log Z_+\big)r^2S
}}
Using the identities in section 2.1, one can show that these
differential equations are compatible with the recursion relation
\eqn\recurrencerew{\eqalign{
 &r(\mu,q-1)^2S(x,\mu,q-1) = \Big(2\sqrt{2}ix-\sqrt{2}\partial_\mu\log
  Z_-(\mu,q)\Big)+S(x,\mu,q)^{-1}\cr
  }}
In other words, one can show that starting from a solution
$S(x,\mu,q_0)$ of \diffeqS\ at some $q=q_0$, the functions
$S(x,\mu,q)$ obtained through \recurrencerew\ also satisfy
\diffeqS\ at $q$.

The next step is to introduce the function $h=h(x,\mu,q)$ which
satisfies
\eqn\hofqdiffeqrew{\eqalign{
 &\partial_\mu \log h = -ix+{1\over\sqrt{2}}r^2 S \cr
 &\partial_x \log h = -i(4x^2+\mu)+(2\sqrt{2}x
 Z_++i\sqrt{2}Z_+')Z_- S-ir^2
}}
One can check that
\eqn\hRrel{
S(x,\mu,q) = {h(x,\mu,q-1)\over h(x,\mu,q)}
}
and that
\eqn\tildePsi{
\Psi = h(x,\mu,q)\left(\matrix{ 1 \cr Z_-(\mu,q)
S(x,\mu,q)}\right)
}
satisfies $Q\Psi = x\Psi$ and $P\Psi=-\partial_x \Psi$ for all
$q$. This completes the proof of \psiidentityagain.

\listrefs

\end